\documentclass[prb,aps,twocolumn,superscriptaddress,showpacs]{revtex4}
\usepackage{graphicx}
\usepackage{amsmath}
\usepackage{color}
\usepackage{bm}
\newcommand{\hc}{\hat{c}^{\phantom{\dagger}}}
\newcommand{\hcd}{\hat{c}^{\dagger}}
\newcommand{\hd}{\hat{d}^{\phantom{\dagger}}}
\newcommand{\hdd}{\hat{d}^{\dagger}}
\newcommand{\ket}[1]{\left| #1 \right \rangle } 
\newcommand{\ve}[1]{{\bm #1 }}
\newcommand{\rmi}{{\rm i}}

\renewcommand{\today}{Version of June 15, 2016, approved by JB, FG} 

\begin{document} 

\title{Interplay of Coulomb interaction and spin-orbit coupling}

\author{J\"org B\"unemann}
\affiliation{Fachbereich Physik, Philipps-Universit\"at Marburg,
  D-35032 Marburg, Germany}
\affiliation{Institut f\"ur Physik, BTU Cottbus-Senftenberg, D-03013 Cottbus,  Germany}
\author{Thorben Linneweber} 
\affiliation{Lehrstuhl f\"ur Theoretische Physik II,
Technische Universit\"at Dortmund, D-44227 Dortmund, Germany}
\author{Ute L\"ow} 
\affiliation{Lehrstuhl f\"ur Theoretische Physik II,
Technische Universit\"at Dortmund, D-44227 Dortmund, Germany}
\author{Frithjof B. Anders} 
\affiliation{Lehrstuhl f\"ur Theoretische Physik II,
Technische Universit\"at Dortmund, D-44227 Dortmund, Germany}
\author{Florian Gebhard}
\affiliation{Fachbereich Physik, Philipps-Universit\"at Marburg,
D-35032 Marburg, Germany}

 \date{\today}
 
\begin{abstract}%
We employ the Gutzwiller variational approach to
investigate the interplay of Coulomb interaction and spin-orbit coupling
in a three-orbital Hubbard model. Already in the 
paramagnetic phase we find a substantial renormalization of the 
spin-orbit coupling that enters the effective single-par\-ticle Hamiltonian
for the quasi-particles. Only close to
half band-filling and for sizable Coulomb interaction
we observe clear signatures of Hund's atomic rules for
spin, orbital, and total angular momentum.
For a finite local Hund's-rule exchange interaction we find
a ferromagnetically ordered state. The spin-orbit coupling
considerably reduces the size of the ordered moment,
it generates a small ordered orbital moment, and it induces
a magnetic anisotropy. To investigate the magnetic anisotropy energy,
we use an external magnetic field that tilts the magnetic moment away 
from the easy axis $(1,1,1)$.
\end{abstract} 
 
\maketitle 

\section{Introduction}
\label{sec_intro}

In atomic physics, the spin-orbit coupling (SOC)
plays an important role because it determines the value of the
total angular momentum in the ground state according to Hund's third rule.
After maximizing the total spin $s$ (first rule)
and the total orbital moment $l$ (second rule), 
the quantum number for the total angular momentum
is  $j=|l-s|$ ($j=l+s$) below (above)
half filling (third rule).~\cite{hund1927a,hund1927b}
The third rule applies in the limit where the SOC is small compared 
to the average Coulomb interaction of the electrons, i.e., for all 
`light atoms', including transition metals. Note that the quantum numbers  
$s$ and $l$ are, in fact, well defined only in the limit of a vanishing SOC.    

For atoms in a solid, the situation is obviously much more complicated 
because neither of the three quantum numbers $s$, $l$, or $j$ 
is well defined due to a breaking of the rotational symmetry. 
Yet, we know that some of the basic mechanisms 
of Hund's rules are still relevant. For example, the maximization 
of the spin is a direct consequence of intra-atomic exchange correlations that 
are caused by the electronic Coulomb interactions. The very
same Coulomb interaction is the reason for magnetic 
order in solids, e.g., in ferromagnets. The  SOC
is only a small perturbation to the dominant Coulomb interaction
in transition metals and their compounds. 
Nevertheless, it can have profound consequences, e.g., for
the direction of the magnetic moment, the so-called `easy axis'.

{}From a theoretician's point of view, the analysis
of the interplay and/or competition of a strong 
local Coulomb interaction and a (comparatively small) SOC
in a solid is rather demanding.
Even the study of simplifying models for the Coulomb
interaction, such as multi-orbital Hubbard models,
poses a tremendously difficult task.
Any study of such models is possible with a limited numerical accuracy
only, e.g., in determining the ground-state energy. Given the 
fundamental uncertainties in the treatment of the sizable
Coulomb correlations it is non-trivial to come to firm
conclusions on the effects of the SOC. Therefore, most theoretical 
studies on the interplay of Coulomb interaction and SOC
focused on insulating or spin states and/or  
assumed a rather large 
SOC.~\cite{PhysRevB.77.195123,refId0,0295-5075-101-2-27003,PhysRevB.89.035112}  
 For the study of (itinerant) $4d$, $5d$ or $f$ electron systems, the dynamical 
 mean field theory  has been used frequently in recent years, see, e.g., 
Refs.~[\onlinecite{PhysRevLett.115.156401,
PhysRevLett.116.106402,PhysRevLett.113.177003,shorikov}]. In such systems,
 however,  
 the SOC tends to be significantly larger than in transition metals and 
 their compounds that we have 
 primarily in mind in our present model study.

In this work, we employ the Gutzwiller 
approach~\cite{buenemann1998} to investigate approximate variational
ground states for multi-orbital Hubbard models.
The analytical evaluation of 
expectation values for Gutzwiller wave functions
poses a difficult many-body problem that requires
additional approximations. Most often applied in the context of multi-band
models is the  `Gutzwiller approximation' which becomes exact for 
the Gutzwiller wave functions
in the limit of infinite spatial
dimensions.~\cite{metzner1987,gebhard1990,buenemann1998,buenemannzumuenster2016}
%
%
It can be used  to evaluate expectation values
for a large set of model parameters, see Sect.~\ref{sec:GWF}.
This allows us to study systematically the subtle interplay of 
Coulomb correlations and spin-orbit coupling.

We consider a Hubbard model with three degenerate $t_{2{\rm g}}$ 
orbitals on a three-dimensional cubic lattice. 
In the first part of our investigation we concentrate on the 
interplay of Coulomb interaction and spin-orbit coupling  
for paramagnetic metallic ground states.
We find that the Coulomb interaction enhances the effective SOC
between the quasi-particles. In addition, we 
investigate the significance of Hund's rules. 
Only Hund's first rule approximately applies
in strongly correlated paramagnetic metallic systems. 

It is well known that for a finite (local) exchange interaction,
multi-orbital Hubbard models tend to favor ferromagnetic states
for sufficiently large Coulomb interactions.
In the second part of our investigation we investigate
if and to what extent the ferromagnetic states are 
modified by the spin-orbit coupling.
We find that the SOC opposes the formation of ferromagnetic order in metals.
While, in the absence of SOC,
the ordered moment has no preferred direction, the SOC aligns it 
along the `easy-axis', and induces a small ordered orbital moment.
 
Recently, the Gutzwiller method and the
density functional theory (DFT) were combined in a self-consistent 
manner;~\cite{ho2008,deng2008} a formal derivation can be found in
Ref.~[\onlinecite{1367-2630-16-9-093034}].
The Gutzwiller-DFT was applied to 
a number of materials, for example to nickel and iron,
see Refs.~[\onlinecite{1367-2630-16-9-093034,ironpaper}],
and references therein. From a methodological point of view,
our model study in this work provides a first step
towards a self-consistent treatment of the SOC
within the Gutzwiller-DFT scheme.

This work is organized as follows.
In Sect.~\ref{sec_method}
we introduce our model and summarize the Gutzwiller variational approach.
In Sect.~\ref{sec_results} we discuss our results
for paramagnetic and ferromagnetic ground states.
Summary and conclusions, Sect.~\ref{sec_conclusion},
close our presentation. Technical details 
are deferred to two appendices.

\section{Models and method}
\label{sec_method}

In this section, we introduce our model and explain  
the Gutzwiller variational approach that we use for its 
investigation.

\subsection{Hamiltonian}

We study a Hubbard model with three 
$t_{2{\rm g}}$ orbitals per site 
on a simple-cubic lattice in three dimensions.  The 
Hamiltonian of this system has the form  
\begin{equation}
\hat{H}=\hat{H}_{0}+\hat{H}_{\rm C}+\hat{H}_{\rm so}\;,
\label{4789}
\end{equation}
where $\hat{H}_0$ denotes the kinetic energy of the electrons,
$\hat{H}_{\rm C}$ describes their Coulomb interaction, and
$\hat{H}_{\rm so}$ models the spin-orbit coupling.

\subsubsection{Kinetic energy and density of states}

We consider electrons that move between $t_{2{\rm g}}$ orbitals $b$ and $b'$
on sites~$i$ and~$j$ of our simple-cubic lattice with~$L$ sites. 
In second quantization the single-particle Hamiltonian reads
\begin{equation}
\hat{H}_{0}=\sum_{i \neq j} \sum_{\sigma,\sigma'}
t^{\sigma,\sigma'}_{i,j} \hcd_{i,\sigma}\hc_{j,\sigma'}\;,
\label{4789b}
\end{equation}
where we introduce the combined spin-orbital index 
\begin{equation}
\sigma\equiv (b,s)\; , \;b\in \{1,2,3\}\;, \; 
s\in\{\uparrow, \downarrow\} \;.
\end{equation}
The crystal-field energies are set to zero,
$t_{i,i}^ {\sigma,\sigma'}=0$.

We use the standard parameterization for the hopping 
amplitudes in~(\ref{4789b}) with some generic Slater-Koster 
parameters~\cite{slater1954}
\begin{eqnarray}
t^{(1),(2),(3)}_{\pi}&=&0.3t,-0.1t,0.025t\;,\label{eq:defSKparametersPM}\\
t^{(2),(3)}_{\sigma}&=&0.1t,0.01t\;,\\
t^{(1),(2),(3)}_{\delta}&=&0.1t,-0.025t,0.02t
\end{eqnarray}
for the electron transfers up to 3rd nearest neighbors. By including 
 hoppings beyond the nearest neighbors we make sure that there are 
no artificial features in  our band structure, such as nesting vectors
 or particle-hole symmetry.  
 In transition metal compounds, the value of $t$ is of the order 
 of $1\,$eV. In our pure model study in this work, we will simply set $t=1$ as 
 our energy unit.

\begin{figure}[t]
\begin{center} 
\includegraphics[width=8.6cm]{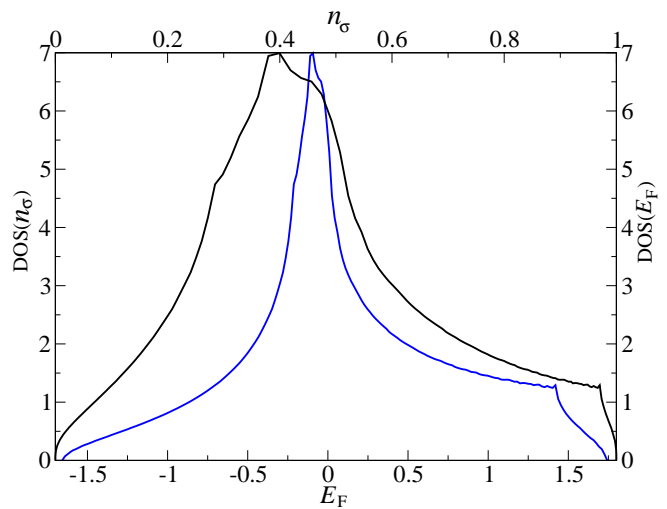} 
\end{center}
\caption{Density of states at the Fermi energy  $E_{\rm F}$  
 as a function of $E_{\rm F}$ (blue) 
 and the orbital occupation $n_{\sigma}$ (black).
\label{Fig:fig1}}
\end{figure}

The single-particle Hamiltonian~(\ref{4789b}) can be readily 
diagonalized in momentum space,
\begin{equation}
\hat{H}_{0}=\sum_{\ve{k}} \sum_{\sigma,\sigma'}
\varepsilon_{\ve{k};\sigma,\sigma'}
\hcd_{\ve{k},\sigma}\hc_{\ve{k},\sigma'}
\label{4789bmomentum}
\end{equation}
with the bare dispersion
\begin{equation}
\varepsilon_{\ve{k};\sigma,\sigma'}
\equiv\frac{1}{L}\sum_{i\neq j}
t^{\sigma,\sigma'}_{i,j} e^{\rmi \ve{k}(\ve{R}_{i}-\ve{R}_{j}) }\;,
\label{eq:baredispersion}
\end{equation} 
and $\ve{k}$ from the first Brillouin zone.
The remaining task is the diagonalization of the
$6\times6$ matrix $\varepsilon_{\ve{k};\sigma,\sigma'}$ for each
$\ve{k}$ to obtain the (bare) band structure.
For non-interacting electrons, all energy levels up to the
Fermi energy~$E_{\rm F}$ are filled in the ground state.
The corresponding density of states at the Fermi-energy $E_{\rm F}$ 
is shown in Fig.~\ref{Fig:fig1} 
as a function of both $E_{\rm F}$ and of the average orbital 
occupation $0\leq n_{\sigma}\leq 1$. The total bandwidth is~$W\approx 3.4$.

Apparently, the Hamiltonian for the kinetic energy
is not particle-hole symmetric, as can be seen from the
density of states at the Fermi energy. Fig.~\ref{Fig:fig1} clearly shows that
${\rm DOS}(n_{\sigma})\neq{\rm DOS}(1-n_{\sigma})$.
To study the influence of the spin-orbit coupling,
we shall later investigate a particle-hole symmetric kinetic energy.
For this case, we use the somewhat artificial Slater-Koster
parameters for electron transfers between nearest neighbors only, 
\begin{equation}
t^{\prime(1)}_{\pi}=0.2\quad,\quad t^{\prime(1)}_{\delta}=0.1\;,
\label{eq:defSKparameterssymmPM}
\end{equation}
which lead to a symmetric density of states of bandwidth~$W'=2$.

In our ferromagnetic calculations we focus on 
the filling $n_{\sigma}\approx 0.4$ where the (paramagnetic) 
density of states has a maximum at the Fermi energy. At such a maximum 
we can expect a stronger tendency towards ferromagnetic order
according to the Stoner criterion.~\cite{Stoner}

\subsubsection{Local interactions}
\label{sec:localinteractions}

The Coulomb and spin-orbit interaction are assumed to be purely local,
\begin{equation}
  \hat{H}_{\rm C}=\sum_i\hat{H}_{i;{\rm C}}\quad , \quad
  \hat{H}_{\rm so}=\sum_i\hat{H}_{i;{\rm so}}\;.
\end{equation}  
The local Coulomb interaction for a model with three degenerate 
 $t_{2{\rm g}}$ orbitals reads~\cite{sugano1970}
\begin{eqnarray}
2\hat{H}_{i;{\rm C}}&=&
U \sum_{b,s}\hat{n}_{i,b,s}\hat{n}_{i,b,\bar{s}}
+\!\!\!\sum_{\genfrac{}{}{0pt}{1}{b(\neq)b'}{s,s'}}\!(U'-\delta_{s,s'}J)
\hat{n}_{i,b,s}\hat{n}_{i,b',s'}\nonumber\\
&&+J\sum_{b(\neq)b'}
\bigg[
\left(
\hcd_{i,b,\uparrow}\hcd_{i,b,\downarrow}
\hc_{i,b',\downarrow}\hc_{i,b',\uparrow}+{\rm h.c.}
\right)\nonumber \\
&&\hphantom{+J\sum_{b(\neq)b'}\bigg[\biggl(}
+\sum_{s}\hcd_{i,b,s}\hcd_{i,b',\bar{s}}
\hc_{i,b,\bar{s}}\hc_{i,b',s}\bigg]\; ,
\label{app3.5b}
\end{eqnarray}
where we use the convention $\bar{\uparrow}=\;\downarrow$, 
$\bar{\downarrow}=\;\uparrow$, and
$\hat{n}_{i,b,s}=\hcd_{i,b,s}\hc_{i,b,s}$ counts the electrons
with spin~$s$ in orbital~$b$ on site $i$.
Note that for  $t_{2{\rm g}}$-orbitals
the three parameters in~(\ref{app3.5b}) are not independent because they 
obey the symmetry relation $U'=U-2J$.~\cite{sugano1970}

For the SOC we use
\begin{equation}
\hat{H}_{i;{\rm so}}=\sum_{\sigma,\sigma'}
\epsilon^{\rm so}_{i;\sigma,\sigma'}\hcd_{i,\sigma}\hc_{i,\sigma'} \; .
\label{456}
\end{equation}
When we work with the following order for our local basis $|\sigma \rangle$, 
\begin{equation}\label{xcb}
|1\rangle=|yz,\uparrow\rangle, \;|2\rangle=|yz,\downarrow\rangle, \;
|3\rangle=|xz,\uparrow\rangle,\ldots,|6\rangle=|xy,\downarrow\rangle,
\end{equation}
the six-dimensional SOC matrix $\tilde{\epsilon}^{\rm so}$ in~(\ref{456})
has the well-known form
\begin{equation}
\tilde{\epsilon}^{\rm so}=-{\rm i} \frac{\zeta}{2}
\left(
\begin{array}{ccc}
0&  - \tilde{\sigma}_3 & \tilde{\sigma}_2 \\
   \tilde{\sigma}_3  &0&- \tilde{\sigma}_1\\
  \tilde{\sigma}_2 & \tilde{\sigma}_1 &0\\
\end{array}
\right)\equiv \zeta \tilde{\Sigma}
\label{aau}
\end{equation}
with the standard two-dimensional Pauli matrices $\tilde{\sigma}_1$,
$\tilde{\sigma}_2$, $\tilde{\sigma}_3$, and the SOC constant $\zeta$. 

The local Hamiltonian 
\begin{equation}\
\hat{H}_{i;{\rm loc}}=\hat{H}_{i;{\rm C}}+\hat{H}_{i;{\rm so}}
\end{equation}
in the $64$-dimensional local Hilbert space is readily diagonalized,
\begin{equation}
\hat{H}_{i;{\rm loc}}|\Gamma \rangle_i =E_{\Gamma}|\Gamma \rangle_i \;.
\end{equation}
For parameter values that are typical for transition metals,
$\zeta/J= 0.2\ldots 1.0$ and $J/U= 0.2$,
the atomic spectrum has a generic form. In table~\ref{tableone} we list the 
degenerate eigenspaces of $\hat{H}_{i;{\rm loc}}$,
ordered by increasing energy for given particle number $0\leq n_{\rm loc}\leq 6$.
We give the degeneracy $g$ of each level, its
total spin~$s$, orbital moment~$l$, and total `angular momentum'~$j$.

\begin{table}[ht]
\centering
\begin{tabular}{c}
\begin{tabular}{l r}
\begin{tabular}[t]{|c|c|c|c|c|c|}
\hline
\# & $n_{\rm loc}$ & $g$ & $s$ &$l$& $j$  \\
\hline \hline  
1 &0  & 1 &0  &0& 0  \\
\hline \hline  
1& 1 & 4 &1/2  &1& 3/2  \\
\hline
2 & 1 & 2 & 1/2 &1& 1/2  \\
\hline\hline  
1 & 2 & 5 & 1 &1& 2  \\
\hline
2 & 2 & 3 &1  &1& 1  \\
\hline
3 & 2 & 1 & 1 &1& 0  \\
\hline
4 & 2 & 5 &0  &2& 2  \\
\hline
5& 2 & 1 & 0 &0&  0 \\
\hline
\end{tabular}
&
\begin{tabular}[t]{|c|c|c|c|c|c|}
\hline
\# & $n_{\rm loc}$ & $g$ & $s$ &$l$& $j$  \\
\hline \hline
1 &6  & 1 &0  &0& 0  \\
\hline  \hline  
1& 5 & 2 &1/2  &1& 1/2  \\
\hline
2 & 5 & 4 & 1/2 &1& 3/2  \\
\hline\hline  
1 & 4 & 1 &1  &1& 0  \\
\hline
2 & 4 &3  & 1 &1&  1 \\
\hline
3 &4  & 5 &1  &1& 2  \\
\hline
4 &4  & 5 &0  &2& 2  \\
\hline
5 &4  & 1 & 0 &0&  0 \\
\hline
\end{tabular}
\end{tabular}\\
\begin{tabular}[t]{|c|c|c|c|c|c|}
\hline
\# & $n_{\rm loc}$ & $g$ & $s$ &$l$& $j$  \\
 \hline \hline
1 & 3 & 4 & 3/2 &0& 3/2  \\
\hline
2 &3  & 4 & 1/2 &2& 3/2  \\
\hline
3 & 3 & 6 &1/2  &2& 5/2  \\
\hline
4 & 3 & 2 &1/2  &1& 1/2  \\
\hline
5 & 3 & 4 &1/2  &1& 3/2  \\
\hline
\end{tabular}
\end{tabular}
\caption{Degenerate eigenspaces of $\hat{H}_{i;{\rm loc}}$,
ordered by energy for a given particle number $0\leq n_{\rm loc}\leq 6$
 with a specification of the degeneracy $g$, total spin~$s$, orbital moment~$l$, 
and total `angular momentum'~$j$.
\label{tableone}}
\end{table}

Since the rotational symmetry is broken in our cubic 
environment, the quantum numbers $l$ and $j$ do, in fact,
not label eigenstates of the  total `angular momentum'
operator. It is well known, however,
that in the $t_{2 \rm {g}}$ sub-space we have
\begin{equation}
\tilde{\ve{l}}^2= \sum_{i\in \{x,y,z  \}}\tilde{l}^2_i=2 \openone 
\end{equation}
for the vector $\tilde{\ve{l}}$  of the three matrices 
\begin{eqnarray}
\tilde{l}_x&=&
\left(
\begin{array}{ccc}
0&  0 & 0 \\
   0  &0& \rmi \\
0   & -\rmi &0\\
\end{array}
\right)\; , \nonumber \\
\tilde{l}_y&=&
\left(
\begin{array}{ccc}
0&  0 & -\rmi \\
   0  &0& 0\\
 \rmi  & 0 &0\\
\end{array}
\right)\;,\nonumber\\
\tilde{l}_z&=&
\left(
\begin{array}{ccc}
0& \rmi  & 0 \\
   -\rmi  &0& 0\\
  0 & 0 &0\\
\end{array}
\right)\;.
\end{eqnarray}
Hence, the orbital moment behaves like that of $l=1$ states
(`T-P equivalence'),~\cite{sugano1970} because
$\langle \hat{\ve{l}}^2\rangle=1(1+1)=2$. To be more precise, one 
finds
\begin{equation}\label{1qw}
\tilde{l}_i=-\tilde{l}^{(l=1)}_i
\end{equation}
where on the r.h.s.~we introduced the
representation of the orbital momentum for ($l=1$) $p$-orbitals.
Due to the T-P equivalence we can label the multiplet states~$|\Gamma \rangle$
by a quantum number $j$ that formally corresponds to a total angular momentum 
of $l=1$ orbitals.
Table~\ref{tableone} shows that Hund's rules are still  
valid for the ground states of all particle numbers
if  we make the replacement $l\to-l$ in Hund's third rule,
as a consequence of eq.~(\ref{1qw}). 
In particular, as seen from table~\ref{tableone},
the local spectrum is not particle-hole symmetric. 
As we will show in  Sect.~\ref{sec:effsoc}, 
the particle-hole asymmetry induced by the SOC
is visible in our itinerant three-band lattice model
even when we work with a symmetric density of states.

\subsection{Gutzwiller wave functions and energy functional}
\label{sec:GWF}

\subsubsection{Wave functions}

For the variational investigation of the Hamiltonian~(\ref{4789}) we use
the Gutzwiller wave functions 
\begin{equation}
|\Psi_{\rm G}\rangle=\prod_{i}\hat{P}_{i}|\Psi_0\rangle\;,
\label{1.3}
\end{equation}
where $|\Psi_0\rangle$ is a normalized single-particle product state
 (Slater determinant) and the 
local Gutzwiller correlator is defined as 
\begin{equation}
\hat{P}_{i}=\sum_{\Gamma,\Gamma^{\prime}}\lambda_{i;\Gamma,\Gamma^{\prime}}
|\Gamma \rangle_{i} {}_{i}\langle \Gamma^{\prime} |\equiv
\sum_{\Gamma_{\rm d}}\lambda_{i;\Gamma_{\rm d} } 
| \Gamma_{\rm d} \rangle_{i} {}_{i}\langle \Gamma_{\rm d}   |   \;.
\label{1.4b}
\end{equation}
Here, we introduce the matrix $\tilde{\lambda}_i$ of (complex) 
variational parameters
$\lambda_{i;\Gamma,\Gamma^{\prime}}$
which allows us to optimize the occupation 
and the form of the eigenstates $|\Gamma_{\rm d}\rangle_{i}$ of $\hat{P}_{i}$.

We assume that the matrix $\tilde{\lambda}_i$ is Hermitian which ensures
that the eigenstates  $|\Gamma_{\rm d} \rangle_{i}$ exist and form a 
basis of the local Hilbert space. Without SOC 
it is usually a sensible approximation to work with a diagonal (and hence real)
matrix $\tilde{\lambda}_i$. For a finite SOC, however, 
it is essential to include at least some non-diagonal elements in  
$\tilde{\lambda}_i$. In this work, we will take into account all 
non-diagonal parameters in $\lambda_{i;\Gamma,\Gamma^{\prime}}$ 
 with states $| \Gamma \rangle_i $ and $| \Gamma' \rangle_i $
 that have the same particle number. 

The evaluation of expectations values with respect to the 
wave function~(\ref{1.3}) poses a difficult many-particle problem
that cannot be solved in general. 
As shown in Refs.~[\onlinecite{buenemann1998,buenemann2005}], 
it is possible to derive analytical expressions for the 
variational ground-state energy in the limit of infinite spatial dimensions 
($D\to \infty$).
An application of this energy functional to finite-dimensional systems 
is usually termed `Gutzwiller approximation'. It will also be
used in this work. 
One should keep in mind, however, that the Gutzwiller approximation has its 
limitations, and the study of some phenomena requires an evaluation 
of expectation values in finite dimensions.~\cite{buenemann2012a,buenemann2012b}

Since the energy functional of the Gutzwiller approximation has been 
derived in detail in previous work, we will only
summarize the main results in this section. 
In the following we are only interested  in systems and wave functions
that are translationally invariant. Hence,  we shall drop 
lattice-site indices whenever this does not lead to ambiguities.  

\subsubsection{Constraints}
\label{con1}

As shown in Refs.~[\onlinecite{buenemann1998,buenemann2005}] it is most 
convenient for the evaluation of Gutzwiller wave functions in 
infinite spatial dimensions to impose the following (local) constraints
\begin{eqnarray}
\langle\hat{P}^{\dagger}\hat{P}^{}\rangle_{\Psi_0}-1&\equiv
&g^{\rm c}_1(\tilde{\lambda},\ket{\Psi_0})=0\;,
\label{1.10a}\\
\langle  \hcd_{\sigma} \hat{P}^{\dagger}\hat{P}^{} \
\hc_{\sigma'}\rangle_{\Psi_0}-C_{\sigma',\sigma}&\equiv&
g^{\rm c}_{\sigma,\sigma'}(\tilde{\lambda},\ket{\Psi_0})=0
\label{1.10b}
\end{eqnarray}
for the local correlation operators $\hat{P}\equiv\hat{P}_i$. 
Here, we introduce the local density matrix $\tilde{C}\equiv \tilde{C}_i$
with the elements
\begin{equation}
C_{i;\sigma,\sigma'}=\langle \hcd_{i,\sigma'}\hc_{i,\sigma}   \rangle_{\Psi_0}\;.
\label{xc}
\end{equation}
Note that the order of indices in~(\ref{xc}) has been chosen deliberately
because it slightly simplifies the analytical results in Sect.~\ref{sec:minimini}.

The constraints can be evaluated by means of Wick's theorem; 
explicit expressions are given in Appendix~\ref{app1}. 
In systems with a high symmetry, the matrix $\tilde{C}$  is often 
diagonal, e.g., for $d$ orbitals in a cubic environment.
In such a case, one usually has to take into account 
only the diagonal constraints~(\ref{1.10b}), because 
the l.h.s.~of~(\ref{1.10b}) for $\sigma\neq \sigma'$ is automatically zero 
for all values of $\tilde{\lambda}_{\rm i}$ that are included in the variational 
Ansatz. Here,  the matrix 
$\tilde{C}$ is non-diagonal in our system with a finite SOC. 
 Even if one introduces a local
basis which has a diagonal local density matrix with respect to
$\ket{\Psi_0}$, see Appendix~\ref{app1}, 
one still has to take into account some non-diagonal constraints. 

\subsubsection{Expectation values}
\label{exp}

Each local operator $\hat{O}_i$, e.g., the operator $\hat{H}_{i;{\rm so}}$, 
can be written as
\begin{eqnarray}
\hat{O}_i&=&\sum_{\Gamma,\Gamma'}O_{\Gamma,\Gamma'}\hat{m}_{i;\Gamma,\Gamma'}\;,\\
\hat{m}_{i;\Gamma,\Gamma'}&\equiv& |\Gamma\rangle_{i} {}_{i}\langle \Gamma'|\;.
\end{eqnarray}
In infinite dimensions the expectation value of  $\hat{O}_i$ has the 
form
\begin{equation}
\langle 
\hat{O}_i\rangle_{\Psi_{\rm G}}
=\sum_{\Gamma_1,\Gamma_2,\Gamma_3,\Gamma_4}O_{\Gamma_2,\Gamma_3}
\lambda_{\Gamma_2,\Gamma_1}^{*}\lambda_{\Gamma_3,\Gamma_4}^{}
\langle \hat{m}_{i;\Gamma_1,\Gamma_4}\rangle_{\Psi_0}\;,
\end{equation}
where the remaining expectation values
\begin{equation}\label{457}
m^0_{i;{\Gamma,\Gamma'}}\equiv
\langle \hat{m}_{i;{\Gamma,\Gamma'}}\rangle_{\Psi_0}
\end{equation}
can readily be evaluated using Wicks theorem, see Appendix~\ref{app1}.

The expectation value of a hopping operator in infinite dimensions 
reads ($i\neq j$)
\begin{equation}
  \big\langle
  \hat{c}_{i,\sigma_1}^{\dagger}\hat{c}_{j,\sigma_2}^{\phantom{+}} \big\rangle_{\Psi_{\rm G}}
  =\sum_{\sigma'_1,\sigma'_2}q_{\sigma_1}^{\sigma'_1}\left( q_{\sigma_2}^{\sigma'_2}\right)^{*}
  \big\langle\hat{c}_{i,\sigma'_1}^{\dagger}\hat{c}_{j,\sigma'_2}^{\phantom{+}}
  \big\rangle_{\Psi_{0}}\;,
\label{8.410} 
\end{equation}
where an analytical expression for the (local) renormalization matrix
$q_{\sigma}^{\sigma'}$ is also given  in Appendix~\ref{app1}. 
Note that the matrix $q_{\sigma}^{\sigma'}$ is, in general, neither real nor
Hermitian. Any symmetries among its elements are caused by those 
of the orbital basis states $|\sigma \rangle$ and the form of the Gutzwiller 
wave function.
For example, if we have no SOC and no magnetic or 
orbital order in our degenerate three-band system, the renormalization matrix 
has the simple form $q_{\sigma}^{\sigma'}=\delta_{\sigma,\sigma'}\sqrt{q}$ with only 
one renormalization factor for all orbitals. 

\subsubsection{Structure of the energy functional}

In a translationally invariant system, the expectation values that we 
introduced in the previous section lead to the 
following variational energy functional (per lattice site)
 \begin{eqnarray}
E_{\rm G}\bigl(\tilde{\lambda},\ket{\Psi_0}\bigr)&=&
\sum_{\substack{\sigma_1,\sigma_2 \\ \sigma'_1,\sigma'_2}}
q^{\sigma'_1}_{\sigma_1}\left(q^{\sigma'_2}_{\sigma_2}\right)^*
E_{\sigma_1,\sigma_2,\sigma'_1,\sigma'_2}
\nonumber \\
&&+
\sum_{\Gamma,\Gamma_1,\Gamma_2}E_{\Gamma}
\lambda_{\Gamma,\Gamma_1}^{*}\lambda_{\Gamma,\Gamma_2}^{}
m^0_{\Gamma_1,\Gamma_2}\;.
\label{ap7.1}
\end{eqnarray} 
Here, we introduce the tensor 
\begin{eqnarray}
E_{\sigma_1,\sigma_2,\sigma'_1,\sigma'_2}&\equiv&
\frac{1}{L}\sum_{i\neq j}   t^{\sigma_1,\sigma_2}_{i,j}
\langle  
\hcd_{i,\sigma'_1}\hc_{j,\sigma'_2}
\big \rangle_{\Psi_0}
\nonumber\\
&=& \frac{1}{L}\sum_{\ve{k}}
\varepsilon_{\ve{k};\sigma_1,\sigma_2}
\big \langle  
\hcd_{\ve{k},\sigma'_1}\hc_{\ve{k},\sigma'_2}
\big \rangle_{\Psi_0}
\label{ap7.2}
\end{eqnarray} 
with the bare dispersion $\varepsilon_{\ve{k};\sigma,\sigma'}$ from
eq.~(\ref{eq:baredispersion}).

The energy~(\ref{ap7.1}) is a function of $\lambda_{\Gamma,\Gamma'}$
and $\ket{\Psi_0}$ where $\ket{\Psi_0}$ enters~(\ref{ap7.1}), (\ref{ap7.2})  
solely through the (non-interacting) density matrix $\tilde{\rho}$
with the elements 
\begin{equation}
\rho_{(i\sigma),(j\sigma')}\equiv
\langle \hat{c}_{j,\sigma'}^{\dagger}
\hat{c}_{i,\sigma}^{\phantom{+}}\rangle_{\Psi_0}\;.
\end{equation}
Note that the non-local elements of $\tilde{\rho}$ ($i\neq j$) determine 
the tensor~(\ref{ap7.2}) while its local elements
\begin{equation}\label{iat}
\rho_{(i\sigma),(i\sigma')}=C_{i;\sigma,\sigma'}\;,
\end{equation}
as introduced in eq.~(\ref{xc}), enter the elements~$q_{\sigma}^{\sigma'}$
of the renormalization matrix,
the expectation value~(\ref{457}),
and the constraints~(\ref{1.10a}), (\ref{1.10b}).
 
The energy 
\begin{equation}\label{459}
E_{\rm G}=E_{\rm G}(\tilde{\lambda},\tilde{\rho},\tilde{C})
\end{equation} 
has to be minimized with respect to the 
variational parameters $\lambda_{\Gamma,\Gamma'}$ and the 
density matrices $\tilde{\rho}$ and $\tilde{C}$  obeying the 
constraints~(\ref{1.10a}), (\ref{1.10b}),  (\ref{iat}), and 
\begin{equation}
\label{16} 
\tilde{\rho}^2=\tilde{\rho}\;.
\end{equation} 
This additional constraint ensures that $\tilde{\rho}$
corresponds to a Slater determinant $|\Psi_0\rangle$. Note that 
introducing the local density matrix $\tilde{C}$ as an 
independent variational object in~(\ref{459}), at the expense of the additional 
constraint~(\ref{iat}), is actually not necessary. Instead one could consider 
the energy solely as a function of $\tilde{\lambda}, \tilde{\rho}$.
Our form of the energy functional, however, will turn out to be slightly 
more convenient because, in the Gutzwiller approximation,  
$\tilde{\rho}$ enters the energy
in a non-linear way only through its local elements.   

\subsubsection{Minimization of the energy functional}
\label{sec:minimini}

We introduce the real and the imaginary parts of the variational parameters
\begin{equation}
\lambda_{\Gamma,\Gamma'}=\lambda_{\Gamma,\Gamma'}^{({\rm r})}+\rmi 
\lambda_{\Gamma,\Gamma'}^{({\rm i})}\;.
\end{equation} 
Due to the Hermiticity of $\tilde{\lambda}$ we have 
\begin{eqnarray}
\lambda_{\Gamma,\Gamma'}^{({\rm r})}&=&\lambda_{\Gamma',\Gamma}^{({\rm r})}\;,\\
\lambda_{\Gamma,\Gamma'}^{({\rm i})}&=&-\lambda_{\Gamma,\Gamma'}^{({\rm i})}\rightarrow
\lambda_{\Gamma,\Gamma}^{({\rm i})}=0\;,
\end{eqnarray} 
which leads to a number $n_v$  of independent (and real) 
variational parameters 
$\lambda_{\Gamma',\Gamma}^{({\rm r/i})}$. They will be considered as 
the components $v_z$ of the $n_v$-dimensional vector
\begin{equation}
\ve{v}=(v_1,\ldots,v_{n_v})^{\rm T}\;.
\end{equation}
The (in general) complex 
constraints~(\ref{1.10a}), (\ref{1.10b}) are not all independent, e.g.,  
because  of the Hermiticity of $\tilde{g}^{\rm c}$. 
We denote the set of all independent real and imaginary parts 
of~(\ref{1.10a}), (\ref{1.10b}) by the $n_c$ real constraints
\begin{equation}\label{iuy}
g_l(\ve{v},\tilde{C})=0\qquad (l=1,\ldots,n_c)\;.
\end{equation}
The constraints~(\ref{iat}), (\ref{16}), and~(\ref{iuy}) 
are implemented via Lagrange parameters $\eta_{\sigma,\sigma'}$, 
$\Omega_{(i\sigma),(j\sigma')}$, and $\Lambda_l$.
This leads to the 
Lagrange functional
\begin{eqnarray}
L_{\rm G}&\equiv&
E_{\rm G}(\ve{v},\tilde{\rho},\tilde{C})
-\sum_l\Lambda_lg_{l}(\ve{v},\tilde{C})\nonumber \\
&&-\sum_{\sigma,\sigma'}\eta_{\sigma,\sigma'}\sum_i
(C_{\sigma',\sigma}-\rho_{(i\sigma'),(i\sigma)})\nonumber\\
&&-\sum_{i,j}\sum_{\sigma,\sigma'}\Omega_{(i\sigma),(j\sigma')}
[\tilde{\rho}^2-\tilde{\rho}]_{(j\sigma'),(i\sigma)}
\label{sgh}
\end{eqnarray}
which provides the basis of our minimization.

As shown, e.g., in Ref.~[\onlinecite{buenemann2012c}],
the minimization of~(\ref{sgh}) 
with respect to $\tilde{\rho}$
leads to the effective single-particle Hamiltonian
\begin{equation}\label{tzs}
\hat{H}^{\rm eff}_0=\sum_{i,j}\sum_{\sigma,\sigma'}(\bar{t}^{\sigma,\sigma'}_{i,j}
+\delta_{i,j}\eta_{\sigma,\sigma'})\hcd_{i,\sigma}\hc_{j,\sigma'}
\end{equation}
with the renormalized hopping parameters
\begin{equation}
\bar{t}^{\sigma_1,\sigma_2}_{i,j}(\ve{v},\tilde{C})=\sum_{\sigma'_1,\sigma'_2}
q^{\sigma_1}_{\sigma'_1}(\ve{v},\tilde{C})
\left(q^{\sigma_2}_{\sigma'_2}(\ve{v},\tilde{C})\right)^*
t^{\sigma'_1,\sigma'_2}_{i,j}\;.
\label{4.7b}
\end{equation}
The optimal Slater determinant $\ket{\Psi_0}$ is the ground state of 
$\hat{H}^{\rm eff}_0$,
\begin{equation}
\hat{H}^{\rm eff}_0\ket{\Psi_0}=E_0^{\rm eff}\ket{\Psi_0}\;.
\label{sdfj00}
\end{equation}
{}From the minimization of~(\ref{sgh}) 
with respect to $\tilde{C}$ we obtain an equation for 
$\eta_{\sigma,\sigma'}$  in~(\ref{tzs}), 
\begin{equation}
\eta_{\sigma,\sigma'}=\frac{\partial}{\partial C_{\sigma,\sigma'}}E_{\rm G}-
\sum_l\Lambda_l\frac{\partial}{\partial C_{\sigma,\sigma'}}g_{l}\;.
\label{sdfj}
\end{equation} 
Finally, the minimization with respect to $\ve{v}$ 
\begin{equation}
\frac{\partial}{\partial v_{Z}}E_{\rm G}
-\sum_l\Lambda_l\frac{\partial}{\partial v_{Z}} g_{l}=0\;
\label{syt}
\end{equation} 
determines the Lagrange parameters $\Lambda_l$ and the optimal value 
of $\ve{v}$.
Equations~(\ref{tzs})--(\ref{syt}) need to be solved self-consistently. 
In Appendix~\ref{app2} we explain in more detail how we solve this 
problem numerically. Note that our minimization algorithm does not require 
 the constraints $g_{l}(\ve{v},\tilde{C})$ to be independent. This is 
 a major advantage over the method that we had proposed in the earlier 
 work~[\onlinecite{buenemann2012c}].

\section{Results}
\label{sec_results}

In the following we discuss the paramagnetic and the ferromagnetic cases
separately. 

\subsection{Paramagnetic ground states}

\subsubsection{Effective spin-orbit coupling}
\label{sec:effsoc}

Without any breaking of spin or orbital symmetries, the minimization 
of the Gutzwiller energy functional leads to effective on-site
energies~(\ref{sdfj}) that have the same form as the SOC~(\ref{aau})
but with the coupling constant $\zeta$ replaced  by $\zeta^{\rm eff}$.
This change from the bare to an effective  coupling constant 
also changes the quasi-particle dispersion of
$\hat{H}_0^{\rm eff}$. Therefore, the energy splittings 
at certain high-symmetry points as seen in ARPES experiments 
are a measure for the effective, not the bare spin-orbit coupling. 
Note that extracting the quasi-particle dispersion from our Gutzwiller approach 
relies on a Fermi-liquid interpretation.~\cite{buenemann2003b}  
However, all changes of the effective single-particle Hamiltonian, e.g., energy shifts,  
are related to changes of certain ground-state expectation values. 
Since the latter are variationally controlled, it is very likely that the exact 
single-particle spectrum reflects the same trends.

\begin{figure}[ht] 
\includegraphics[width=8cm]{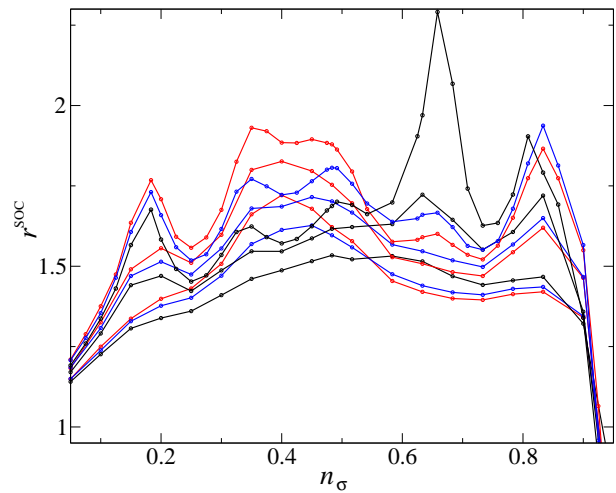} 
\caption{SOC-renormalization $r^{\rm SOC}$  as a function of the 
orbital occupation $n_{\sigma}$ for  $J=0$, $\zeta=0.05$ (red),  
 $\zeta=0.1$ (blue), $\zeta=0.2$ (black) and  $U=2,3,4$ (in ascending order).
\label{Fig:fig2}}
\end{figure} 

\begin{figure}[t] 
\includegraphics[width=8cm]{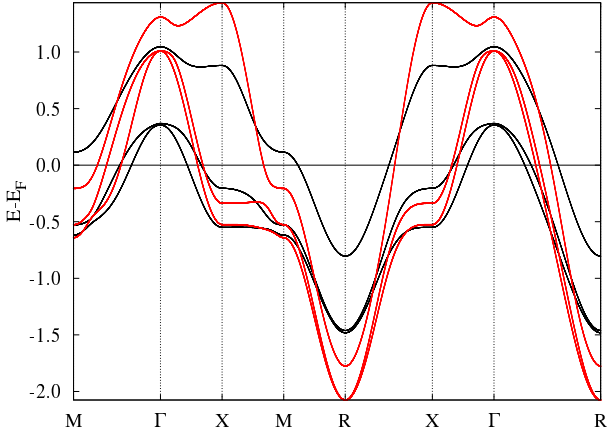} 
\caption{Quasi-particle bands along high-symmetry lines
for $\zeta=0.2$, $n_{\sigma}=2/3$, $U=0$, $J=0$ (red), and    
$U=4$, $J/U=0.2$ (black). } 
\label{fig:bands}
\end{figure}

In Fig.~\ref{Fig:fig2} we show the renormalization  of $\zeta$,
\begin{equation}
r^{\rm SOC}\equiv \zeta^{\rm eff}/\zeta
\end{equation}
as a function of the orbital occupation~$n_{\sigma}$ for the three bare 
values $\zeta=0.05, 0.1, 0.2$ and interaction 
parameters $U=2,3,4$ and $J=0$. Apparently,
the effective spin-orbit coupling increases as a function of~$U$,
apart from a small region of an almost filled shell
where $\zeta^{\rm eff}(U)<\zeta$. 
For $U=4$, which is approximately equal to the band width, 
the spin-orbit couping can be renormalized by a factor two or more,
$r^{\rm SOC}(U=4,n_{\sigma}=2/3,\zeta=0.2)\approx 2.3$. 
This substantial increase is clearly visible in the 
quasi-particle band structure, see Fig.~\ref{fig:bands},
where we show the bare ($U=J=0$) and renormalized band structures
($U=4,J/U=0.2$) for $n_{\sigma}=2/3$ and $\zeta=0.2$.
For example, the splitting of the bands at the $\Gamma$-point and the R-point
is noticeably enhanced in presence of the Coulomb interaction.

\begin{figure}[b] 
\includegraphics[width=8cm]{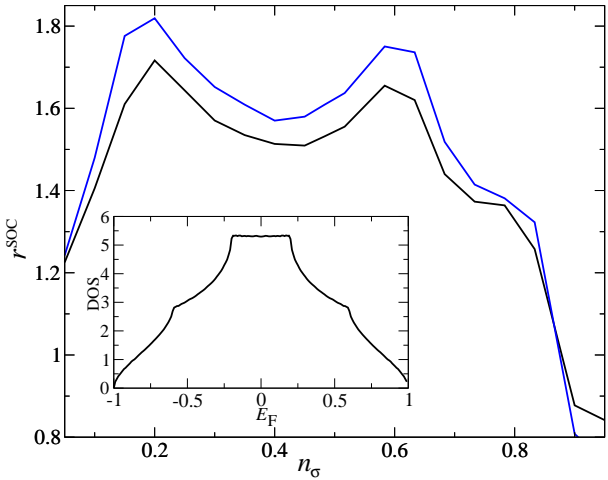} 
\caption{SOC-renormalization $r^{\rm SOC}$  
for the symmetric density of states from the nearest-neighbor 
electron transfers~(\ref{eq:defSKparameterssymmPM}), as a function of the 
orbital occupation $n_{\sigma}$ for  $\zeta=0.1$, $J=0$,     
 $U=3$ (black) and $U=4$ (blue); inset: density of states 
at the Fermi energy.\label{Fig:fig3}}
\end{figure} 

The renormalization $r^{\rm SOC}$ 
is not monotonous as a function of the bare coupling~$\zeta$.
Moreover, it is {\em not\/} particle-hole symmetric, 
i.e., it is not invariant under the transformation $n_{\sigma}\to 1-n_{\sigma}$. 
This is only partly caused by the particle-hole asymmetry
of the bare density of states 
in Fig.~\ref{Fig:fig1}.  As discussed already in
Sect.~\ref{sec:localinteractions},
the SOC inherently breaks the particle-hole symmetry. To illustrate this
point, we show the results for a symmetric density of states that
results from the nearest-neighbor 
electron transfers~(\ref{eq:defSKparameterssymmPM}),
displayed in the inset of Fig.~\ref{Fig:fig3}. As seen from the figure,
the SOC alone induces a particle-hole 
asymmetry in the renormalization of the effective  spin-orbit coupling.
We not in passing, that band structures with a fairly similar density of states
may, nevertheless, display a very different $n_{\sigma}$ dependence of
$r^{\rm SOC}$. Apparently, the full momentum dependence 
of the band structure determines the details of the $r^{\rm SOC}$-curves.

\begin{figure}[t] 
\includegraphics[width=8cm]{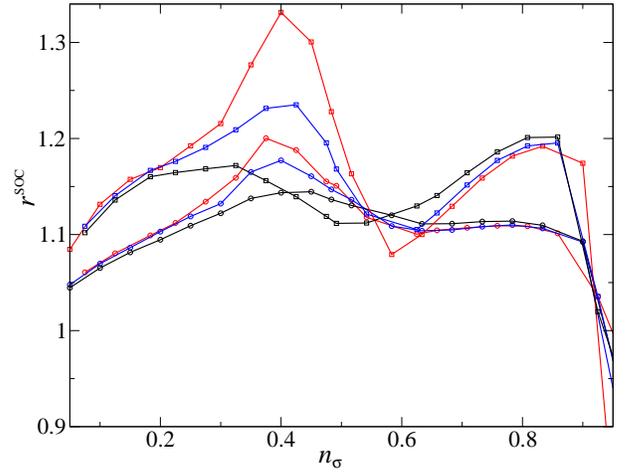} 
\caption{SOC-renormalization $r^{\rm SOC}$ as a function of the 
orbital occupation $n_{\sigma}$ for  $J/U=0.2$, $\zeta=0.05$ (red),  
 $\zeta=0.1$ (blue), $\zeta=0.2$ (black), and  $U=1$ (circles), 
$U=2$ (squares).
\label{Fig:fig5}}
\end{figure} 

\begin{figure}[t] 
\includegraphics[width=8cm]{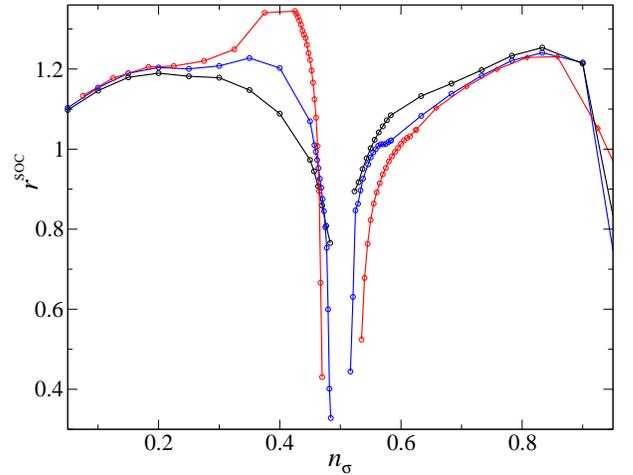} 
\caption{SOC-renormalization $r^{\rm SOC}$ as a function of the 
orbital occupation $n_{\sigma}$ for  $J/U=0.2$, $\zeta=0.05$ (red),  
 $\zeta=0.1$ (blue), $\zeta=0.2$ (black), and $U=2.5$.
\label{Fig:fig5b}}
\end{figure} 

For finite values of the exchange interaction $J$, the  
effective  coupling constants are smaller than for $J=0$, in general. 
This can be seen in Figs.~\ref{Fig:fig5} and~\ref{Fig:fig5b}
where we show the renormalization  for $J/U=0.2$,
and  $U=1,2$ (Fig.~\ref{Fig:fig5}), $U=2.5$ (Fig.~\ref{Fig:fig5b}). 
Note that for $U=U_{\rm c}\lesssim 2.5$ there appears a 
Brinkmann-Rice type of insulating phase\cite{brinkmann1970} at half filling
where the renormalization matrix $\tilde{q}$ is zero.  
Therefore we could perform our calculations
shown in Fig.~\ref{Fig:fig5b} only away from half filling.  
  
The dependence of the renormalization on the band-filling $n_{\sigma}$
appears to be even more complicated for finite $J$, 
in particular in the region around half filling. One must keep in mind, 
however, that there is a `trivial' contribution to the renormalization 
of $\zeta$ which simply stems from
the band-width renormalization induced by the 
renormalization matrix $q^{\sigma'}_{\sigma}$. To understand this effect,
we consider, for the sake of argument, a renormalization matrix of the 
simplest form $q^{\sigma'}_{\sigma}=\delta_{\sigma,\sigma'}\sqrt{q}$.
In that case, the 
effective hopping parameters in~(\ref{tzs}) are given by
$\bar{t}^{\sigma,\sigma'}_{i,j}=q t^{\sigma,\sigma'}_{i,j}$. Hence, in order 
to obtain the same expectation values of  $\ket{\Psi_0}$ as in the 
non-interacting limit, we must introduce a scaling 
$\zeta\to q \zeta< \zeta $. The effect of the enhancement of $\zeta^{\rm eff}$
is therefore amplified by the renormalization of the hopping 
parameters. 

\begin{figure}[hb] 
\includegraphics[width=8cm]{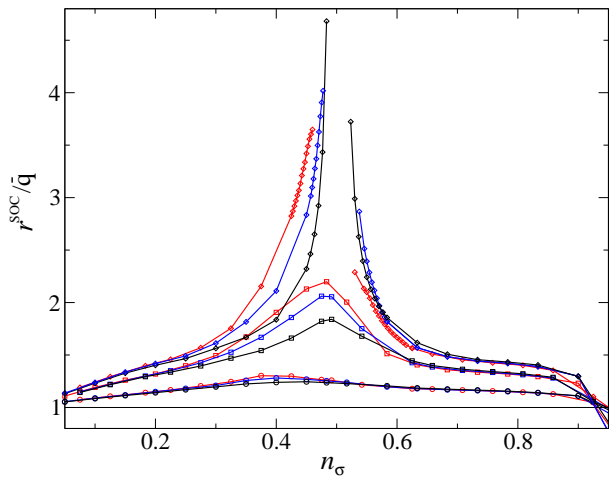} 
\caption{Relative SOC-renormalization $r^{\rm SOC}/\bar{q} $ as a 
function of the 
orbital occupation $n_{\sigma}$ for  $J/U=0.2$, $\zeta=0.05$ (red),  
$\zeta=0.1$ (blue), $\zeta=0.2$ (black) and  $\zeta=0.1$ (blue),
$\zeta=0.2$ (black), and  $U=1$ (circles), 
$U=2$ (squares), $U=2.5$ (diamonds).
\label{Fig:fig6}}
\end{figure}

For a more quantitative analysis, we define an average value $\bar{q}$
of the bandwidth renormalization through
\begin{equation}
\bar{q}=\langle \hat{H}_0\rangle_{\rm G}/ \langle\hat{H}_{0}\rangle_0\;,
\end{equation}
i.e., $\bar{q}$ quantifies the reduction of the average kinetic energy
in presence of the Coulomb interaction.
The relative SOC-renormalization is then plotted in Fig.~\ref{Fig:fig6} 
for the same parameters as in Fig.~\ref{Fig:fig5}. It shows that the
non-trivial renormalization is, in fact, largest in the region around
half filling. 
Moreover, it is actually fairly independent of the 
bare SOC, a feature that cannot be seen in the original 
representation of the data in Fig.~\ref{Fig:fig5}. 

\subsubsection{Hund's rules in a solid?}
\label{sec:hundsrule}

In the introduction we raised the question
if, and to what extent, Hund's rules are still discernible in a solid. 
To clarify this issue, we define the three `quantum numbers' $s,l,j$ via the 
local expectation values 
\begin{eqnarray}
\langle \hat{\ve{S}}_i^{2} \rangle_{\rm G}&=&s(s+1)\;,\nonumber \\
\langle \hat{\ve{L}}_i^{2} \rangle_{\rm G}&=&l(l+1)\;,\nonumber \\
\langle (\hat{\ve{S}}_i+\hat{\ve{L}}_i)^{2} \rangle_{\rm G}&=&j(j+1)\;.
\end{eqnarray}%

\begin{figure}[ht] 
\includegraphics[width=8cm]{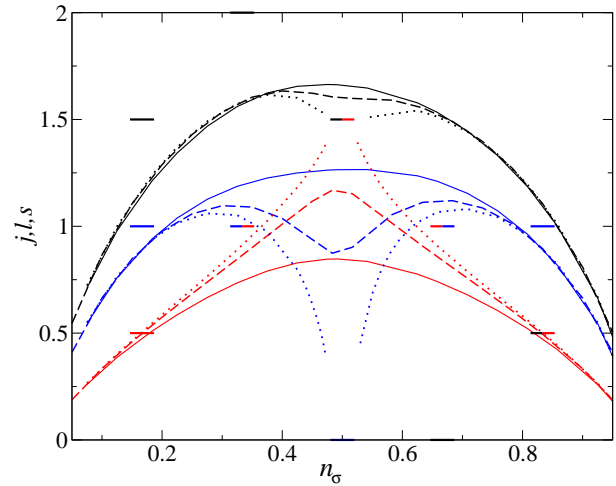} 
\caption{Quantum numbers $j$ (black), $l$ (blue), $s$ (red) as a 
function of the orbital occupation $n_{\sigma}$ for  $J/U=0.2$, $\zeta=0.05$,
and  $U=1$ (solid), $U=2$ (dashed), $U=2.5$ (dotted).
\label{Fig:fig7}}
\end{figure}

Figure~\ref{Fig:fig7} shows these three numbers for 
$\zeta=0.05$, $J/U=0.2$ and $U=1,2,2.5$.  The bars give the values in the 
atomic limit, as extracted from the ground states in 
table~\ref{tableone}. As expected, all quantum numbers move 
towards their atomic values when we increase the Coulomb interaction
parameters. This is best visible near half-filling when
the system is close to the metal-insulator transition that 
appears at half filling. 

\begin{figure}[b] 
\includegraphics[width=8cm]{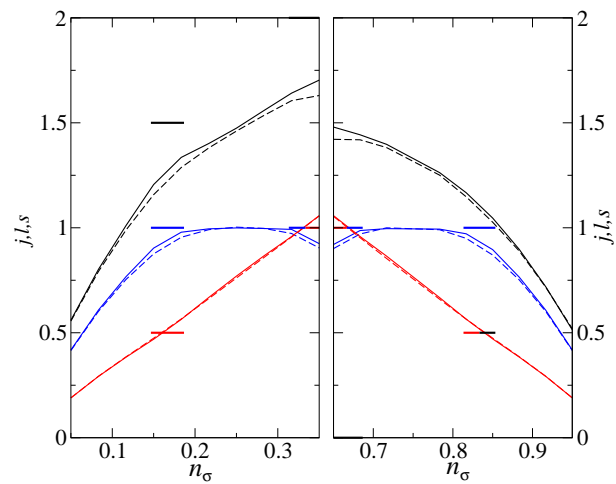} 
\caption{Quantum numbers $j$ (black), $l$ (blue), $s$ (red)
as a function of the 
orbital occupation $n_{\sigma}$ for  $J/U=0.2$, $\zeta=0.05$,
and  $U=6$ (solid), $U=9$ (dashed).
\label{Fig:fig7b}}
\end{figure} 

As shown in previous work,\cite{buenemann1997,buenemann1998}
this transition is of 
first-order where 
in the Gutzwiller insulating state all atoms are in their ground state. 
This means that at $U_c$
all three quantum numbers will jump to their atomic values at half filling. 
For all other (integer) fillings, the system is still rather itinerant 
and some of the quantum numbers, in particular $j$, deviate significantly 
from their atomic values. This is best visible at a filling of 
$n_{\sigma}=2/3$ where the value of~$j$ is far off its atomic value 
$j_{\rm atomic}=0$. The results change only slightly when we increase 
the values of $U$ (and $J$) as can be seen from Fig.~\ref{Fig:fig7b} where 
we display $j$, $l$, and $s$ for larger values of $U$ away from half filling.

The difference between the behavior close to half filling and
the other integer fillings can be understood from the
atomic spectra. The high-spin ground state at half filling is only 
slightly changed by a small SOC and, most importantly, 
its degeneracy is not lifted. 
Hence, the energy difference between the Hund's-rule ground state and 
the first excited state is of the order of $J$. In contrast, at all other
integer fillings, the ground states are created by a splitting
of the (degenerate) ground states at $\zeta=0$, caused by the SOC.
Therefore, the energy difference between the Hund's-rule ground state and 
the first excited states
is much smaller away from half filling. As a consequence, 
it is energetically not favorable to lose a lot of kinetic energy by only 
occupying the Hund's-rule ground state.  
Unlike in the half-filled case, the Hund's-rule ground state
does not dominate the quantum numbers 
in the metallic phase at or around other integer fillings.
As seen from Figs.~\ref{Fig:fig7} and~\ref{Fig:fig7b},
only Hund's first rule is seen to be obeyed in strongly correlated
paramagnetic metals close to integer fillings.
 
\subsection{Ferromagnetic ground states}

Without the spin-orbit coupling, the Hamiltonian commutes with the total spin
operator. Hence, the energy of a ferromagnetic ground state cannot depend 
on the direction of the magnetic moment. For finite SOC,
there is a preferred direction of the moment, the so-called `easy axis'. 
In order to find this axis, we minimize the energy 
functional with respect to $|\Psi_0\rangle$ without any bias 
on the magnetic-moment direction using a completely general 
matrix $\eta_{\sigma,\sigma'}$. It turns out
that in our system and for the parameters considered in this 
section, the magnetic moment always points into the $(1,1,1)$-direction.   

\begin{figure}[t] 
\includegraphics[width=8cm]{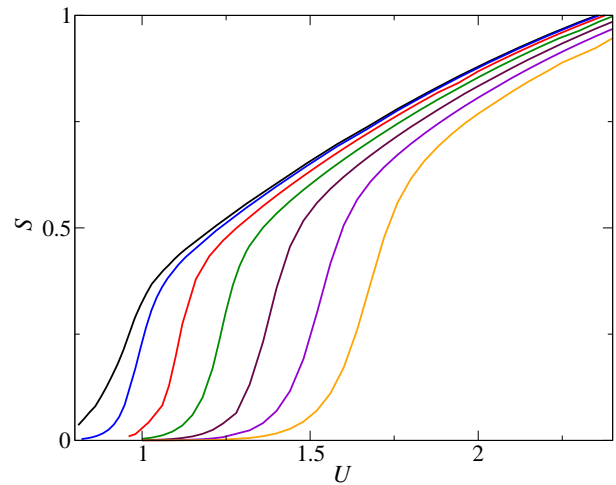}
\caption{Spin $S$ in $(1,1,1)$-direction as a function of $U$ with $J/U=0.2$ 
for $n_{\sigma}=0.4$ and 
 $\zeta=0.05$ (blue),  $0.1$ (red),  $0.15$ (green),  $0.2$ (maroon),  
$0.255$ (violet),  $0.3$ (orange).\label{Fig:S}}
\end{figure} 

\subsubsection{Ordered moment}

In Fig.~\ref{Fig:S} we display the total spin
$S\equiv |\langle \hat{\ve{S}}_i  \rangle|$
for seven different values of $\zeta$ ($ 0\leq\zeta\leq  0.3)$ as a function 
of $U$ for $J/U=0.2$. 
As seen from the figure,
the SOC destabilizes the ferromagnetic order, i.e.,
the value $U_c$ for noticeable ferromagnetic order ($S>0.1$)
substantially increases as a function of~$\zeta$.
Concomitantly, the ordered magnetic moment $m=2S$ strongly depends
on the SOC as long as the magnetic order is weak, $S<1/2$.
The SOC becomes a small perturbation only in the saturation region, $S>1$.

\begin{figure}[b] 
\includegraphics[width=8cm]{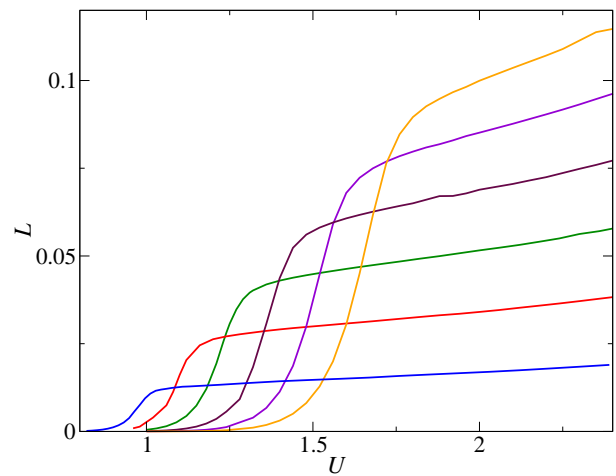} 
\caption{Orbital moment $L$ in (1,1,1)-direction as a function of
$U$ with $J/U=0.2$ for $n_{\sigma}=0.4$ and $\zeta=0.05$ (blue),  $0.1$ (red),
$0.15$ (green),  $0.2$ (maroon),     
$0.255$ (violet),  $0.3$ (orange).\label{Fig:L}}
\end{figure} 

The SOC not only reduces the ordered spin moment,
it also induces an orbital moment, i.e.,
$L\equiv |\langle \hat{\ve{L}}_i  \rangle|$ is non-zero. 
This is shown in Fig.~\ref{Fig:L} where we display $L$ for the 
same parameters as in Fig.~\ref{Fig:S}, apart from  $\zeta=0$ where $L=0$. 
The orbital contribution to the magnetic moment, however, remains rather small,
of the order of 10\% of the spin moment, 
especially for values of $\zeta<0.1$ that are realistic for transition
metals. Therefore, the gain in orbital moment does not compensate
the loss in the ordered spin moment induced by the spin-orbit coupling.

\begin{figure*}[t] 
\includegraphics[width=15cm]{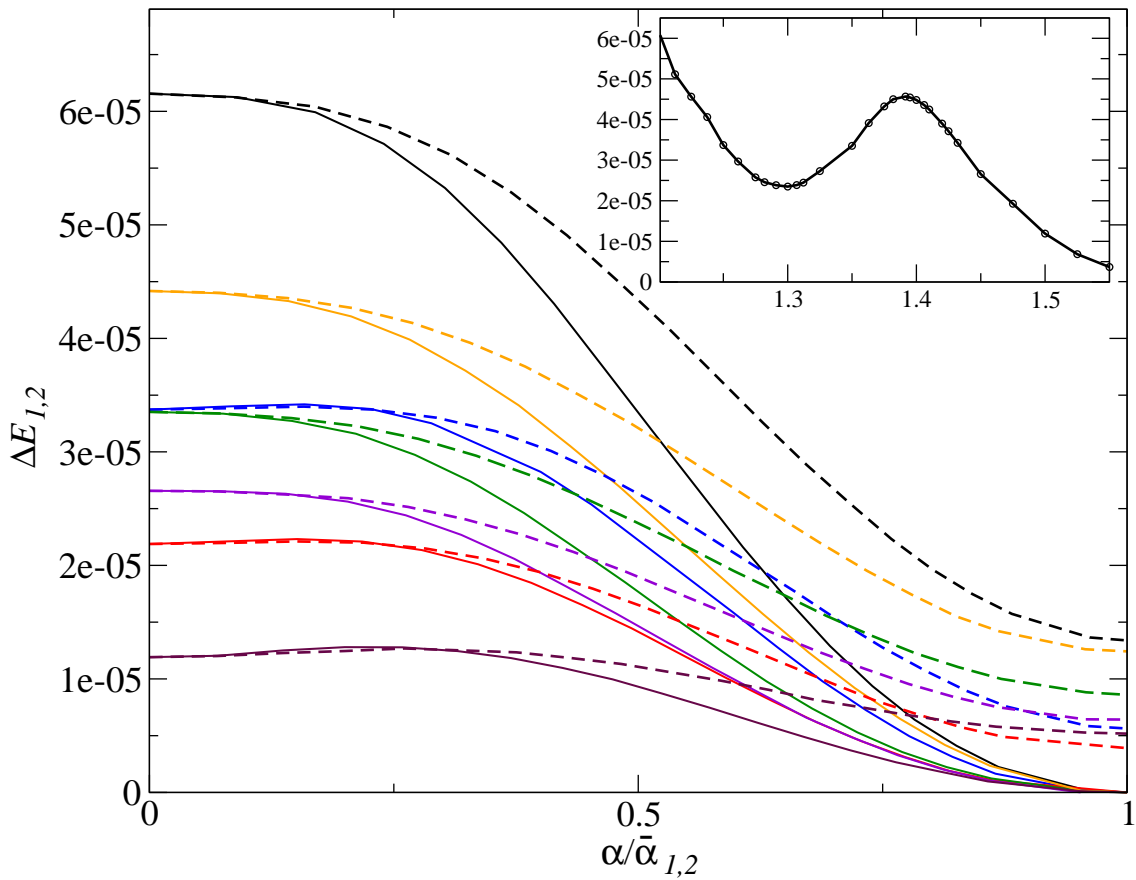} 
\caption{Anisotropy energy $\Delta E_{1,2}$
as a function of the magnetic-moment direction 
that is rotated, (1), from  $(1,0,0)$  to  $(1,1,1)$ (solid lines) and, (2),
from
$(1,0,0)$  to  $(1,1,0)$ (dashed lines) with maximum rotation angles
$\bar{\alpha}_1=\arccos{1/\sqrt{3}}$ and   $\bar{\alpha}_2=\pi/4$; 
parameters: 
$U=1.2$ (black), $U=1.25$ (blue), $U=1.3$ (red), $U=1.35$ (green), 
$U=1.4$ (orange), $U=1.45$ (violet),  $U=1.5$ (maroon), $J/U=0.2$, $\zeta=0.1$;
inset: maximal anisotropy energy as a function of $U$ for  $J/U=0.2$,
$\zeta=0.1$. }
\label{plot20}
\end{figure*}

\subsubsection{Anisotropy energy}

Finally, we take a look at the `anisotropy energy', i.e., the 
dependence 
of the energy on the magnetic-moment direction.  
To this end, we could introduce additional constraints 
that fix the moment direction during the minimization. 
However, this would require additional 
programming work that we prefer to avoid. Therefore, we 
apply an external magnetic field that allows us to change the 
magnetic-moment direction. In fact, this is how the
anisotropy energy would actually be measured.

Since our field is just a 
technical tool, we couple it to the spin only, i.e., we add 
\begin{equation}
\hat{H}_{B}=-B \sum_i \ve{e}_B\cdot \ve{S}_i 
\end{equation}
to the Hamiltonian of our system. Here, $\ve{e}_B$ is 
the direction of the magnetic field that we adjust in our calculations. 
The size of the field amplitude $B$ must be chosen with care
to obtain meaningful results. On the one hand, 
it must be large enough to force the magnetic moment into all 
directions that we aim to investigate, i.e., in the ground state
we must approximately find $\langle \ve{S}_i  \rangle_{\rm G}||\ve{e}_B$.
On the other hand, the variation in the field contribution 
to the energy must be small compared 
to the variation of the system's energy that we actually want to 
determine. Meeting these criteria becomes difficult, in particular, 
in the region of small magnetic moments.  In all calculations that we are
going to present 
below, we found that a  field amplitude of $B= 0.002$ leads to meaningful 
results for the anisotropy energy.

In the following we consider rotations of the magnetic moment from 
the $(1,0,0)$ direction, (1), into the $(1,1,1)$ direction, and, (2),
into the $(1,1,0)$
direction. The corresponding maximal rotation angles are 
$\bar{\alpha}_1=\arccos{(1/\sqrt{3})}$ and   $\bar{\alpha}_2=\pi/4$, 
respectively. From our minimization we obtain the two energies $E_{1,2}(\alpha)$
as a function of the angle $\alpha$. Since, as mentioned before, the easy
axis always points into the $(1,1,1)$-direction, 
we define the anisotropy energy
as $\Delta E_i(\alpha)\equiv E_i(\alpha)-E_1(\bar{\alpha}_1)$.
This quantity 
is displayed in Fig.~\ref{plot20} for several values of $U$ 
(and consequently also different values
of the magnetic moment) for $J/U=0.2$ and $\zeta=0.1$.  The figure shows that,
although the anisotropy energy is quite small,
of the order of several ten $\mu$eV per site, our approach is perfectly 
capable to resolve it.  The maximal anisotropy energy $\Delta E_{\rm max}$, 
i.e., its value for $\alpha=0$ is a non-trivial function of~$U$. 
This can be seen from the inset 
of Fig.~\ref{plot20} where we display $\Delta E_{\rm max}$.

\begin{figure}[t] 
\includegraphics[width=8cm]{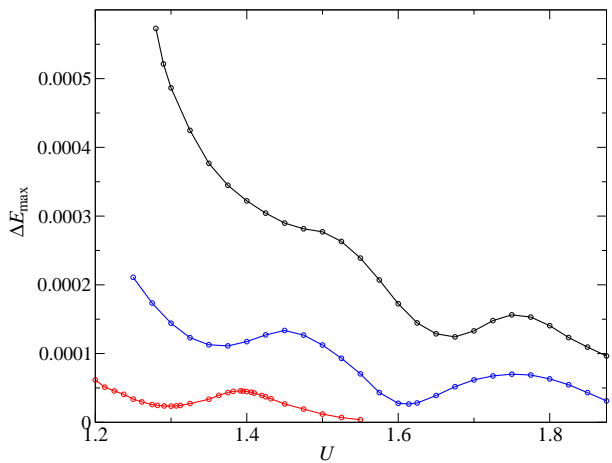} 
\caption{Maximal anisotropy energy as a function of $U$ for  $J/U=0.2$, 
$\zeta=0.2$ (black), $\zeta=0.15$ (blue), $\zeta=0.1$ (red).}
\label{plot30}
\end{figure}

When we increase the SOC, the anisotropy energies change 
significantly, see Fig.~\ref{plot30} where 
we show $\Delta E_{\rm max}$ as a function of~$U$ for  
$\zeta=0.1, 0.15, 0.2$. The non-monotonic  
behavior of $\Delta E_{\rm max}$  has its 
cause in the band structure. For example, the maxima 
in the red and blue curves and the corresponding structure 
in the black curve correspond to almost the same 
magnetization, cf.~Fig.~\ref{Fig:S}. 

\begin{figure}[ht] 
\includegraphics[width=8cm]{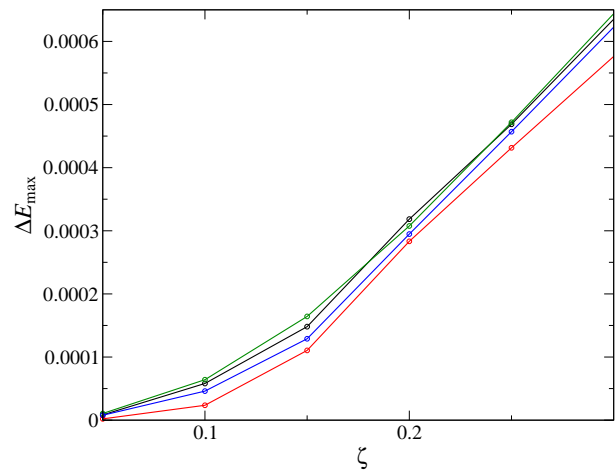} 
\caption{Maximal anisotropy energy as a function of $\zeta$ for values
 of $U$ with the same  ordered spin moment of $S=0.4$ (black),  
 $S=0.45$ (blue), $S=0.5$ (red),  $S=0.6$ (green),  and  $J/U=0.2$. }
\label{plot40}
\end{figure}

To extract the genuine 
$\zeta$ dependence of $\Delta E_{\rm max}$, it is best to 
consider states with the same moment. This is done in 
Fig.~\ref{plot40} where we display $\Delta E_{\rm max}$  
as a function of $\zeta$ for values
of $U$ which lead to the same  ordered spin moments. These curves
reveal that the anisotropy depends very sensitively 
on $\zeta$ for small values of~$\zeta$  
whereas it becomes linear for sizable $\zeta$.  

\section{Summary and conclusions}
\label{sec_conclusion}

In this work we investigated the interplay of local Coulomb interactions
and the spin-orbit coupling in a three-orbital Hubbard model
in three dimensions. 
Based on the Gutzwiller approximation to
general multi-band Gutzwiller wave functions,
we find that the Coulomb interaction leads to a considerable
renormalization of the effective SOC in paramagnetic metals;
the spin-orbit couplings can be enhanced over their atomic values
by a factor of more than two. This effect could be seen in experiment 
as enhanced band splittings in the quasi-particle dispersion. 

Hund's rules determine spin and orbital moments of an atom.
In metallic systems, signatures of Hund's rules are visible only close to
half band-filling. For all other (integer) fillings, 
the local Hund'-rule ground states cannot dominate over states
with other quantum numbers because this would be very unfavorable
for the electrons' kinetic energy. At best, 
Hund's first rule applies in strongly correlated metallic systems
close to integer fillings.

For ferromagnetic ground states, we find magnetization curves that are 
significantly influenced by the spin-orbit coupling.
Overall, the SOC tends to destabilize the ferromagnetic order. 
For example, it shifts the onset of ferromagnetism to higher values
of the Coulomb parameters. In the presence of an ordered spin moment,
the SOC has two main effects: (i), the magnetic spin moment points into 
a preferred direction (easy axis), and, (ii), it generates 
a small but finite orbital moment in the same 
direction as the spin moment.

We analyzed the magnetic anisotropy by applying 
an external magnetic field with constant strength
and varying direction. Our method is capable to 
resolve the anisotropy energy which can be 
rather small for spin-orbit couplings 
that are realistic for transition metals. As a function of the 
Coulomb interaction, the  anisotropy energy shows a 
non-monotonic behavior which we could trace back to 
details of the electronic band structure. 

In this study we worked with the most general Ansatz for a Gutzwiller
wave function. For the calculation of anisotropy energies,
it is mandatory to avoid
the often used approximation of a diagonal variational-parameter matrix
because this approximation results in anisotropy energies that can be off 
by several orders of magnitude.

For our three-band model, it is possible to include
all elements of the variational-parameter matrix.
Of course, this cannot be done for five $d$-bands. Therefore,
strategies must be developed to include only the most significant
matrix elements. In a separate, more technical work, we analyze in detail
the importance of non-diagonal variational parameters, 
and show how
to obtain accurate results with a properly chosen subset of such parameters.~\cite{dd} 

Our method can directly be applied to materials that can be described by effective three-band models, e,g,, 
Sr$_2$RuO$_4$.  It will be interesting to see the consequences of the substantial 
spin-orbit coupling on the ground-state phase diagram and other electronic properties of these 
systems.\cite{PhysRevLett.115.247001,PhysRevLett.116.106402}

\section*{Acknowledgements}
We thank R.~Schade for valuable discussions on 
optimization algorithms. 

This work was supported in part by the Priority Programme 1458
of the Deutsche Forschungsgemeinschaft (DFG)
under GE 746/10-1. 
T.L., U.L., and F.B.A.\ acknowledge the financial support by the Deutsche
For\-schungs\-ge\-mein\-schaft and the Russian Foundation of
Basic Research through the Transregio TRR~160.

The authors gratefully acknowledge the computing time granted 
by the John-von-Neumann Institute for Computing (NIC),
and provided on the supercomputer JURECA at J\"ulich Supercomputing Centre (JSC)
under project no.\ HDO08.  
\appendix

\section{Energy functional and its derivatives}
\label{app1}

\subsection{Local basis}

The local density matrix~(\ref{xc}) 
is non-diagonal when we include the spin-orbit coupling.
For a fixed state $|\Psi_0 \rangle$, however, 
we can always find a local basis, described by operators 
\begin{equation}\label{88s}
\hdd_{i,\gamma}=\sum_{\sigma}u_{i;\sigma,\gamma}\hcd_{i,\sigma}, \quad, \quad
\hd_{i,\gamma}=\sum_{\sigma}u^{*}_{i;\sigma,\gamma}\hc_{i,\sigma}
\end{equation}
and a unitary matrix $\tilde{u}_{i}$, so that the local density matrix $\tilde{D}_i$
is diagonal, 
\begin{equation}\label{99s}
D_{i;\gamma',\gamma}\equiv\langle \hdd_{i,\gamma}
\hd_{i,\gamma'}   \rangle_{\Psi_0}=\delta_{\gamma,\gamma'}
n_{i,\gamma}\;.
\end{equation}
Working with this new orbital basis $|\gamma\rangle$ is quite useful 
because the energy functional~(\ref{ap7.1}) as well as the 
constraints~(\ref{1.10a}),  (\ref{1.10b}) have a much simpler 
form, see Sect.~\ref{app:energyfunc}.

In general, the basis
$|\gamma\rangle$ is not uniquely defined. For instance, in our three-band model 
without any charge or magnetic order, our local density matrix has the form
\begin{equation}\label{5f}
\tilde{C}=n_0 \openone -\Delta n_0^{\rm so}\tilde{\Sigma}
\end{equation}
with $\tilde{\Sigma}$ as defined in~(\ref{aau}).   
The diagonalization of~(\ref{5f}) leads to a two-fold and a four-fold
degenerate set of states $|\gamma\rangle$ with the occupation numbers
$n_0-2\Delta n_0^{\rm so}$ and $n_0+\Delta n_0^{\rm so}$, respectively. 
Therefore, the states $|\gamma\rangle$ are defined only 
up to an arbitrary unitary transformation within these two degenerate 
sub-spaces. Even for a system with three non-degenerate orbitals
there would be a remaining two-fold degeneracy
in the spectrum of~$\tilde{C}$.

\subsection{Atomic spectrum}

We introduce the configuration basis $\ket{I}$ of the local Hilbert space, 
\begin{equation}\label{4.30a}
\ket{I}\equiv\prod_{\sigma \in I}\hcd_{\sigma}\ket{0}\equiv
\hcd_{\sigma_1}\dots\hcd_{\sigma_{|I|}}\ket{0}\;,
\end{equation}
where the operators $\hcd_{\sigma}$ are in ascending order, i.e., we have
$\sigma_1<\sigma_2\ldots<\sigma_{|I|}$ where $|I|$ is the number of particles
in state $\ket{I}$. Using the standard mathematical notations for set operators, 
we frequently encounter
the states $\ket{I\cup  \sigma}$ or  $\ket{I\backslash  \sigma}$
which result from the local creation/annihilation of an electron.
Since we work with fermions,
we define the minus-sign function
\begin{equation}
{\rm fsgn}(\sigma,I)\equiv \langle I\cup  \sigma | \hcd_{\sigma} |I  \rangle\;.
\end{equation}
With the basis~(\ref{4.30a}), we can readily set up the local 
 Hamilton matrix
\begin{equation}
H^{\rm loc}_{I,I'}=\langle I| \hat{H}_{\rm loc}|I' \rangle
\end{equation}
and determine 
its eigenstates
\begin{equation}\label{4.60a}
|\Gamma\rangle =\sum_{I}T_{I,\Gamma}\ket{I}
\end{equation}
by standard numerical techniques. 
 For the numerical minimization of the Gutzwiller energy functional, 
however, we prefer to work 
 with the orbital states $\ket{\gamma}$ and its corresponding 
 configuration basis
\begin{equation}\label{4.30aa}
\ket{J}\equiv\prod_{\gamma \in J}\hdd_{\gamma}\ket{0}\equiv
\hdd_{\gamma_1}\dots\hdd_{\gamma_{|I|}}\ket{0}\;.
\end{equation}
One way to determine the expansion of $|\Gamma\rangle$ with respect to
 this basis, 
 \begin{equation}\label{4.60c}
|\Gamma\rangle =\sum_{J}A_{J,\Gamma}\ket{J}\;,
\end{equation}
would be to transform the local Hamiltonian $\hat{H}_{\rm loc}$
to the basis  $|\gamma\rangle$ and to set up and diagonalize
the Hamilton matrix $H^{\rm loc}_{J,J'}$. 
Alternatively, one may  determine the eigenstates~(\ref{4.60a}) and 
calculate the coefficients $A_{J,\Gamma}$ in~(\ref{4.60c}) from the formula
 \begin{eqnarray}
A_{J,\Gamma} &=&  \sum_I
T_{I,\Gamma} \langle J | I \rangle \; ,  \nonumber \\[3pt]
\langle J | I \rangle &=& 
\text{Det} (u^*_{\sigma_i,\gamma_j}) 
\quad , \quad (\sigma_i \in I, \gamma_j \in J) \; .
\label{defAbyu}
  \end{eqnarray}

\subsection{Energy functional}
\label{app:energyfunc}

For a (still general) orbital basis $|\gamma\rangle$, we find the following 
 expression for the constraints~(\ref{1.10a}), (\ref{1.10b}),
\begin{eqnarray}\label{5.5dd}
\sum_{\Gamma,\Gamma_1,\Gamma_2}
\lambda_{\Gamma,\Gamma_1}^{*}\lambda_{\Gamma,\Gamma_2}^{}
m^{0}_{\Gamma_1,\Gamma_2}&=&1\;,\\\label{5.5bdd}
\sum_{\Gamma,\Gamma_1,\Gamma_2}
\lambda_{\Gamma,\Gamma_1}^{*}\lambda_{\Gamma,\Gamma_2}^{}
m^{0}_{\Gamma_1\cup \gamma,\Gamma_2\cup \gamma'}
&=&\delta_{\gamma,\gamma'}n_{\gamma}\;,
\end{eqnarray}
where  
\begin{eqnarray}
|\Gamma\cup \gamma \rangle
&\equiv& \hat{d}^{\dagger}_{\gamma}|\Gamma \rangle
 =\sum_{J(\gamma \notin J)}{\rm fsgn}(\gamma,J)A_{J,\Gamma}|J \cup \gamma\rangle\;,
\nonumber \\
&& \label{wet}\\
m^{0}_{\Gamma,\Gamma'}&\equiv&\langle \hat{m}_{\Gamma,\Gamma'}  \rangle_{\Psi_0}
\; .\label{wet2}
\end{eqnarray}
Since $\ket{J}$ is a basis of the local Hilbert space, all expectation values
 of the form~(\ref{wet2}) are determined by the determinants 
\begin{equation}\label{miiprime}
m^{0}_{J,J'}\equiv \langle \hat{m}_{J,J'}  \rangle_{\Psi_0} =\left|
\begin{array}{cc}
\Omega^{J,J'}&-\Omega^{J,\bar{J}}\\
\Omega^{\bar{J},J'}&\bar{\Omega}^{\bar{J},\bar{J}}
\end{array}
\right|\;.
\end{equation}
Here, $\Omega_{J,J'}$ are the matrices
\begin{equation}
\Omega_{J,J'}=\left(
\begin{array}{cccc}
D_{\gamma'_1,\gamma_1}&D_{\gamma'_2,\gamma_1}&\ldots&D_{\gamma'_{|J'|},\gamma_1}\\
D_{\gamma'_1,\gamma_2}&D_{\gamma'_2,\gamma_2}&\ldots&D_{\gamma'_{|J'|},\gamma_2}\\
\vdots&\vdots&\ddots&\vdots\\
D_{\gamma'_1,\gamma_{|J|}}& D_{\gamma'_2,\gamma_{|J|}} &\ldots&D_{\gamma'_{|J'|},\gamma_{|J|}}
\end{array}
\right)\;,
\end{equation}
in which the entries are the elements of the 
uncorrelated local density matrix~(\ref{99s})
that belong to the configurations $J=(\gamma_1,\ldots,\gamma_{|J|})$ and
$J'=(\gamma'_1,\ldots,\gamma'_{|J'|})$. The matrix 
$\bar{\Omega}^{\bar{J},\bar{J}}$ 
in~(\ref{miiprime}) is defined by
\begin{equation}\label{cdf}
\bar{\Omega}_{\bar{J},\bar{J}}=\left(
\begin{array}{cccc}
1-D_{\gamma_1,\gamma_1}&-D_{\gamma_1,\gamma_2}&\ldots&-D_{\gamma_{|\bar{J}|},\gamma_1}\\
-D_{\sigma_2,\sigma_1}&1-D_{\sigma_2,\sigma_2}&\ldots&-D_{\gamma_{|\bar{J}|},\gamma_2}\\
\vdots&\vdots&\ddots&\vdots\\
-D_{\gamma_1,\gamma_{|\bar{J}|}}&-D_{\gamma_2,\gamma_{|\bar{J}|}}&\ldots&
1-D_{\gamma_{|\bar{J}|},\gamma_{|\bar{J}|}}
\end{array}
\right)\;,
\end{equation}
with $\gamma_i\in \bar{J}\equiv (1,\ldots,N)\backslash (J\cup J') $. 

So far we have not used yet the defining condition~(\ref{99s})
of the $|\gamma\rangle$-basis. By applying it, the expectation
values~(\ref{miiprime}) have the much simpler form
\begin{eqnarray}
m^{0}_{J,J'}&=&\delta_{J,J'}m^{0}_{J} \nonumber \; , \\
m^{0}_{J}&=&\prod_{\gamma \in J}n_{\gamma}\prod_{\gamma \notin J}(1-n_{\gamma})  \;.
\label{eq:simplifymJ}
\end{eqnarray}
It is this simplification that makes the use of the $|\gamma\rangle$-basis
particularly convenient in the evaluation of ground-state expectation values. 
For the calculation of derivatives with respect to 
$D_{\gamma,\gamma'}$, however, we have to start from the general
expression~(\ref{miiprime}), see Sect.~\ref{as3}.

With the above results, eqs.~(\ref{wet2})--(\ref{cdf}) ,
we can calculate the local energy as
\begin{equation}
E_{\rm loc}= \sum_{\Gamma,\Gamma_1,\Gamma_2}E_{\Gamma}
\lambda_{\Gamma,\Gamma_1}^{*}\lambda_{\Gamma,\Gamma_2}^{}
m^0_{\Gamma_1,\Gamma_2}\;.
\end{equation}
In the $|\gamma\rangle$-basis we have explicitly
\begin{equation}
m^0_{\Gamma_1,\Gamma_2}= \sum_{J_1,J_2} A_{J_1,\Gamma_1}A_{J_2,\Gamma_2}^*
m^0_{J_1,J_2} \; .
\end{equation}
For a ground-state calculation, this expression can be simplified further 
using eq.~(\ref{eq:simplifymJ}).

Finally, the  renormalization matrix has the form
\begin{eqnarray}
q_{\gamma}^{\gamma'}&=&\sum_{\Gamma_1,\ldots,\Gamma_4}
\lambda^{*}_{\Gamma_2,\Gamma_1}
\lambda^{}_{\Gamma_3,\Gamma_4}\langle \Gamma_2|\hdd_{\gamma}  
|\Gamma_3 \rangle \nonumber\\
&&\hphantom{\sum_{\Gamma_1,\ldots,\Gamma_4}}
\times \sum_{J_1,J_4}A_{J_1,\Gamma_1}A^{*}_{J_4,\Gamma_4}
H^{\gamma'}_{J_1,J_4}
\label{qmat}
\end{eqnarray}
with 
\begin{eqnarray}
H^{\gamma'}_{J_1,J_4}&\equiv&(1-f_{\gamma',J_1})\langle J_4  |
\hd_{\gamma'} |J_4\cup \gamma'  \rangle
m^{0}_{J_1,J_4\cup \gamma'}\nonumber\\
&&+\left(
f_{\gamma',J_4}m^{0}_{J_1\backslash \gamma',J_4}+
(1-f_{\gamma',J_4})m^{0;\gamma'}_{J_1\backslash \gamma',J_4}
\right) \nonumber \\
&&\hphantom{+}\times \langle J_1 \backslash \sigma' |\hd_{\gamma'} |I_1  \rangle
\label{8sgdd}
\;,
\end{eqnarray}
and 
\begin{equation}
f_{\gamma,J}\equiv\langle J |\hdd_{\gamma}\hd_{\gamma} |J  \rangle
\end{equation}
is either zero or unity.
Here, the expectation value $m^{0;\gamma'}_{J_1\backslash \gamma',J_4}$ 
has the same form as the one in~(\ref{miiprime}), except that the index 
$\bar{J}$ 
has to be replaced by $\bar{J} \backslash \gamma'$. 
We need the general result~(\ref{qmat})
for the renormalization matrix 
for the calculation of derivatives with respect to non-diagonal elements 
of $D_{\gamma,\gamma'}$, see Sect.~\ref{as3}.
For a ground-state calculation one can use 
eq.~(\ref{99s}) and obtain the simpler expression
\begin{equation}\label{8.460} 
q_{\gamma}^{\gamma'}=\frac{1}{n_{\sigma'}}
\sum_{\Gamma_1\ldots\Gamma_4}\lambda^{*}_{\Gamma_2,\Gamma_1}
\lambda^{}_{\Gamma_3,\Gamma_4}
\langle \Gamma_2|
\hdd_{\gamma}
|\Gamma_3\rangle
m_{\Gamma_1,\Gamma_4 \cup \gamma'}^0
\end{equation}
which may also be written  in the form~\cite{PhysRevB.76.165110} 
\begin{equation}
q_{\gamma}^{\gamma'}=\frac{1}{n_{\sigma'}}
\langle \hat{P}^{\dagger} \hdd_{\gamma} \hat{P}^{} \hd_{\gamma'}  \rangle_{\Psi_0}\;.
\end{equation}

In summary, the Gutzwiller energy functional in the  
 $|\gamma\rangle$-basis is given as
 \begin{eqnarray}
E_{\rm G}(\ve{v},\tilde{\rho},\tilde{D})&=&
\sum_{\substack{\gamma_1,\gamma_2 \\ \gamma'_1,\gamma'_2}}
q^{\gamma'_1}_{\gamma_1}\left(q^{\gamma'_2}_{\gamma_2}\right)^*
E_{\gamma_1,\gamma_2,\gamma'_1,\gamma'_2}\nonumber\\
&&+
\sum_{\Gamma,\Gamma_1,\Gamma_2}E_{\Gamma}
\lambda_{\Gamma,\Gamma_1}^{*}\lambda_{\Gamma,\Gamma_2}^{}
m^0_{\Gamma_1,\Gamma_2}\;.\nonumber 
\\
&&\label{ap7.1b}
\end{eqnarray} 
Here, we applied the transformation to the $|\gamma\rangle$-basis
\begin{equation}
E_{\gamma_1,\gamma_2,\gamma'_1,\gamma'_2}
=\sum_{\substack{\sigma_1,\sigma_2 \\ \sigma'_1,\sigma'_2}}
u^*_{\sigma_1,\gamma_1}u_{\sigma_2,\gamma_2}u_{\sigma'_1,\gamma'_1}u^*_{\sigma'_2,\gamma'_2}
E_{\sigma_1,\sigma_2,\sigma'_1,\sigma'_2}
\end{equation}
and 
\begin{equation}
q_{\gamma}^{\gamma'}=\sum_{\sigma,\sigma'}u_{\sigma,\gamma}u^*_{\sigma',\gamma'}
q_{\sigma}^{\sigma'}\;.
\end{equation}

\subsection{Derivatives}
\label{as3}

The minimization algorithm which we explain in this section
requires the calculation of derivatives of the energy 
and of the constraints with respect to the variational 
parameters $v_z$ and the local density matrix $\tilde{C}$ or  $\tilde{D}$. 

\subsubsection{Derivatives with respect to $v_Z$}

The constraints, the local energy, and the renormalization factors
are all quadratic functions of the variational 
parameters $v_z$, i.e., they are of the form
\begin{equation}
f(\ve{v})=\sum_{Z,Z'} f_{Z,Z'} v_{Z'}v_{Z}\;.
\end{equation}
The fast calculation of derivatives
\begin{equation}\label{3we}
\partial_{v_{Z}} f(\ve{v})=\sum_{Z'} (f_{Z,Z'}+f_{Z,Z'}) v_{Z'}
\end{equation}
is then possible if 
all coefficients  $f_{Z,Z'}$ are stored in the main memory. 
In our calculations we observe 
that the number of contributing coefficients $f_{Z,Z'}$ in the expansion
is particularly large  in the  renormalization factors
when we include non-diagonal elements 
in the variational parameter matrix $\lambda_{\Gamma,\Gamma'}$. 
Hence, our  minimization for the three-orbital model 
that includes all $n_v=924$  non-diagonal variational parameters 
is numerically much more demanding than the minimization, e.g., 
for a five-orbital model with only diagonal parameters  ($n_v=1024$).  

\subsubsection{Derivatives with respect to $C_{\sigma,\sigma'}$}

For the calculation of the effective on-site energies~(\ref{sdfj}), 
we need to determine the derivatives of the energy and of the constraints
with respect to  $C_{\sigma,\sigma'}$. Again, it is  easier to calculate 
the derivatives first in the $\gamma$-basis and then transform them via
\begin{equation}
\frac{\partial}{\partial C_{\sigma,\sigma'}}=
\sum_{\gamma,\gamma'}u^*_{\sigma,\gamma}u_{\sigma',\gamma'}
\frac{\partial}{\partial D_{\gamma,\gamma'}}\;.
\end{equation}
For the derivatives of the constraints and of the local energy, we just need to 
determine the derivative of~(\ref{miiprime}). This gives
\begin{eqnarray}
\frac{\partial }{\partial D_{\gamma,\gamma}}m^0_{J,J'}=\delta_{J,J'}m^0_{J,J}
\left\{ 
\begin{array}{cl}
1/n_{\gamma}&{\rm for}\;\;  \gamma \in J\\
-1/(1-n_{\gamma})&{\rm for}\;\; \gamma \notin J
\end{array}
 \right.\nonumber \\
\end{eqnarray}
for $\gamma=\gamma'$, and
\begin{equation}
\frac{\partial }{\partial D_{\gamma',\gamma}}m^0_{J,J'}
=\delta_{\bar{I},I \backslash \gamma}
\delta_{\bar{I},I'\backslash \gamma'}\frac{m^0_{\bar{I},\bar{I}}}
{(1-n_{\gamma})(1-n_{\gamma'})}
\end{equation}
for $\gamma\ne \gamma'$, where $\gamma \in J$ and $\gamma' \in J'$.
The only remaining problem is to calculate derivatives of the object 
$m^{0;\bar{\gamma}}_{J,J'}$ that appears in the definition of the renormalization 
matrix, eqs.~(\ref{qmat}), (\ref{8sgdd}) with respect to $D_{\gamma',\gamma}$. 
It contributes only when $\gamma \neq \bar{\gamma}$ and 
$\gamma' \neq \bar{\gamma}$. Then we can use the simple relationship
\begin{equation}
\frac{\partial}{\partial D_{\gamma',\gamma}}m^{0;\bar{\gamma}}_{J,J'}
=\frac{1}{1-n_{\bar{\gamma}}}\frac{\partial}{\partial D_{\gamma',\gamma}}m^0_{J,J'}\;.
\end{equation}

\section{Minimization algorithm}
\label{app2}

\subsection{Inner minimization} 

For a given single-particle state $|\Psi_0 \rangle$, or, equivalently, a 
given single-particle density matrix $\tilde{\rho}$, we have to minimize 
the energy functional~(\ref{459}) obeying the constraints~(\ref{iuy}). In 
Ref.~[\onlinecite{buenemann2012c}]
we introduced a very efficient method for this minimization which was 
used in  a number of previous studies, for example on elementary iron 
and nickel.~\cite{1367-2630-16-9-093034,ironpaper} 
This method, however, is only applicable if  the gradients 
\begin{equation}
\ve{F}^l\equiv \partial_{\ve{v}}g_l(\ve{v})
\end{equation}
of the constraints~(\ref{iuy}) are linearly independent because it 
requires a matrix $W_{l,l'}\equiv \ve{F}^l\cdot  \ve{F}^{l'}$ 
to be regular.

In principle, this problem can be overcome by 
a group-theoretical analysis that identifies 
the maximum set of independent constraints. Such a solution, however, 
is rather cumbersome and it runs into difficulties if one aims
to study the transition between minima with different point-group symmetries.
Even if we ensure that
the gradients  $\ve{F}^l$ are linearly independent, however, we 
observe that the algorithm introduced 
in Ref.~[\onlinecite{buenemann2012c}]
becomes prohibitively slow when we aim to minimize the energy 
functional for a general (complex) variational parameter matrix 
$\lambda_{\Gamma,\Gamma'}$. 

For this reason we tested a couple of alternative 
minimization algorithms that are discussed in textbooks
on numerical optimization.~\cite{num-mini-book}
We found the `Penalty and Augmented Lagrangian Method' (PALM)
to be most useful in our context when combined with an unconstrained
Broyden-Fletcher-Goldfarb-Shanno (BFGS) minimization. 
We shall briefly summarize these methods in the following. 

\subsubsection{PALM} 

In the PALM  one studies the functional 
\begin{equation}
L^{\rm PALM}_{\rm G}(\ve{v},\{\Lambda_{l}\},\mu)\equiv
E_{\rm G}(\ve{v})-\sum_l\Lambda_lg_{l}(\ve{v})
+\frac{\mu}{2}\sum_l[g_{l}(\ve{v})]^2\;,\label{palm}
\end{equation}
which contains Lagrange parameter terms ($\sim\Lambda_l$) and penalty 
terms ($\sim\mu$). In a pure `penalty method' one would set  $\Lambda_l=0$
and minimize~(\ref{palm}) for a given value 
of $\mu>0$. If, in the minimum $\ve{v}=\ve{v}_0$, the constraints 
are sufficiently well fulfilled, i.e., 
\begin{equation}\label{palm2}
\sum_lg_{l}(\ve{v}_0)^2< g_{\rm c}^2
\end{equation}
with some properly chosen value of $g_{\rm c}$, we may consider
$E_0= E_{\rm G}(\ve{v}_0)$ as a decent approximation for 
the Gutzwiller ground-state energy.
Otherwise, we increase $\mu$ and start another minimization.

For our Gutzwiller energy functional it turns out 
that the convergence to the minimum is much faster when we 
use a full PALM algorithm with Lagrange parameters 
$\Lambda_l\neq 0$. This method works as follows.~\cite{num-mini-book}
\begin{itemize}
\item[(i)]
Start from some initial values  
$\Lambda_l=\Lambda_{l;0}$ and $\mu=\mu_0$, e.g., $\Lambda_{l;0}=1$ and 
$\mu_0=50 | E_{\rm G}(\ve{v}^{\rm nc})|$ where $v^{\rm nc}_l$ are the 
variational parameters in the non-interacting limit, i.e., with 
$\lambda_{\Gamma,\Gamma'}=\delta_{\Gamma,\Gamma'}$. 
\item[(ii)]
Minimize 
\begin{equation}
L^{\rm PALM}_{{\rm G};0}(\ve{v})\equiv
L^{\rm PALM}_{\rm G}(\ve{v},\{\Lambda_{l;0} \},\mu_0)
\label{7tr}
\end{equation}
with respect to $\ve{v}$.
For this step we use the method of steepest descent 
combined with the BFGS method, see Sect.~\ref{sec:BFGS}.
We denote the minimum found in step~(ii) by $\ve{v}_0$.
\item[(iii)]
Set 
\begin{eqnarray}
\Lambda_{l;k+1}&=&\Lambda_{l;k}-\mu_{k}g_l(\ve{v}_0)\; ,\\
\mu_{k+1}&=&\beta \mu_{k}
\end{eqnarray}
with some properly chosen number $\beta>1$. In our calculations we worked 
 with  $\beta=2$. 
\item[(iv)] Go back to step~(ii) until eq.~(\ref{palm2}) is satisfied. 
\end{itemize}

\subsubsection{Steepest decent and BFGS method} 
\label{sec:BFGS}

We still have to choose a method for the
unconstrained minimization in step~(ii) in the PALM. It is a major
advantage of our Gutzwiller minimization that calculating
gradients of the energy or of the constraints works just as fast as the 
calculation of these objects themselves. Of course, this is only the case
when we use eq.~(\ref{3we}) and do not try to calculate the gradients
numerically from the  difference quotient. 

Let $E(\ve{v})$ be our functional and
\begin{equation}\label{asd}
\ve{F}_0= \left.\partial_{\ve{v}}E(\ve{v})\right|_{\ve{v}=\ve{v}_0}
\end{equation}
its gradient at the point $\ve{v}_0$. Then the simplest way of 
minimizing  $E(\ve{v})$ is the `method of steepest descent' where
the one-dimensional function 
\begin{equation}
\Delta E(\alpha) =E(\ve{v}_0+\alpha \ve{F}_0)\;
\end{equation}
is minimized with respect to $\alpha$. 
Instead of the optimal value $\alpha=\alpha_0$, in practical numerics we use a
value $\tilde{\alpha}_0$ that reduces the value of our functional $E(\ve{v})$.
We calculate a new point 
$\ve{v}_0 \to \ve{v}_0+\tilde{\alpha}_0 \ve{F}_0 $
and reiterate the procedure until $|\ve{F}_0|$ is below a pre-determined 
threshold. It is the decisive advantage
of this method that it always converges towards a (potentially local) 
minimum as long as the functional is well-behaved,
which we can take for granted in physics.
The main disadvantage of the method is its rather slow 
convergence. Therefore, we found it necessary
to combine it with a faster algorithm, the BFGS method,
which, however, works reliably only in the vicinity of the minimum. 

The starting point of the BFGS method is a second-order expansion
of the functional
\begin{equation}
E(\ve{v}_0+\delta \ve{v})\approx E(\ve{v}_0) +\ve{F}_0\cdot \delta \ve{v}
+ \frac{1}{2} \delta \ve{v}^{\rm T}\cdot \tilde{H}_0\cdot  \delta \ve{v}\; ,
\end{equation}
where $\tilde{H}_0$ is the Hessian matrix of second derivatives at the point 
$\ve{v}_0$. Provided that $\tilde{H}_0$ is positive definite,
the right-hand site is minimized for
\begin{equation}\label{sd}
\delta \ve{v}=-\tilde{B}_0\cdot \ve{F}_0\; ,
\end{equation}
where $\tilde{B}_0=\tilde{H}_0^{-1}$. Making iterative steps in the 
variational parameter space by means of eq.~(\ref{sd}) is a 
multi-dimensional version of the Newton method.

The main obstacle of the Newton method
is the numerical calculation of $\tilde{H}_0$ and the solution of 
eq.~(\ref{sd}). Therefore, it is better to use a so-called 
`quasi Newton method' of which BFGS is one example. This method 
employs eq.~(\ref{sd}) without calculating $\tilde{B}_0$ 
(or $\tilde{H}_0$) exactly. It works as follows.~\cite{num-mini-book}
\begin{itemize}
\item[(i)]
Start at some point $\ve{v}_k$ and calculate
the  gradient $\ve{F}_k$ and the inverse $\tilde{B}_k$ of the Hessian matrix. 
Due to the benign structure of our functional we can afford 
this initial calculation of $\tilde{B}_k$ because it is 
done only once.
\item[(ii)]
Calculate the new point 
\begin{equation}
\ve{v}_{k+1}=\ve{v}_{k}-\tilde{B}_k\cdot \ve{F}_k \; .
\end{equation}
\item[(iii)]
Calculate $\ve{F}_{k+1}$ from
\begin{equation}
\ve{F}_{k+1}= \left.\partial_{\ve{v}}E(\ve{v})\right|_{\ve{v}=\ve{v}_{k+1}}
\end{equation}
and an approximate update of $\tilde{B}_k$ from
\begin{equation}
\tilde{B}_{k+1}=(\tilde{1}-\alpha_k\ve{s}_k\ve{y}^{\rm T}_k)\tilde{B}_{k+1}
(\tilde{1}-\alpha_k\ve{y}_k\ve{s}^{\rm T}_k)+\alpha_k\ve{s}_k\ve{s}^{\rm T}_k
\; ,
\end{equation}
where 
\begin{eqnarray}
\ve{s}_k&\equiv&\ve{v}_{k+1}-\ve{v}_{k}\;,\nonumber\\
\ve{y}_k&\equiv&\ve{F}_{k+1}-\ve{F}_{k}\;,\nonumber\\
\alpha_k&\equiv&\ve{y}^{\rm T}_k\ve{s}_k\;.
\end{eqnarray}
\item[(iv)]
Go back to step~(ii) until $|\ve{F}_{k}|$ is below some pre-defined threshold.
\end{itemize}

Within the BFGS method it is not ensured
that going from $\ve{v}_{k}$ to $\ve{v}_{k+1}$
always leads to a decrease of our functional. Therefore, we need
the method of steepest decent as a backup to 
reach a region in the variational parameter space
where the BFGS method converges.

\subsection{Outer minimization}

Given the optimum variational parameters $\ve{v}_0$
from the inner minimization
we need to determine a new single-particle state by means of 
eqs.~(\ref{tzs})--(\ref{syt}). All derivatives in eqs.~(\ref{sdfj})--(\ref{syt})
are calculated with the formulae given in eq.~(\ref{asd}). 
Then, the remaining problem is the calculation
of the Lagrange parameters $\Lambda_l$ from eqs.~(\ref{syt}).
The number $n_v$ of these linear equations is usually much 
larger than the number of Lagrange parameters $n_{\rm c}$. Due to a possible 
inter-dependence of the constraints,
the solution of the equations may not be unique.
Hence, we cannot use the trick 
of Ref.~[\onlinecite{buenemann2012c}] (see Sec.\ 4.2.1 of that work),
which led to a number of  $n_{\rm c}$ linear equations.

Here, we choose to determine one of the infinitely many possible sets
of Lagrange parameters by minimizing the functional
\begin{equation}
Y(\{\Lambda_l\})= \sum_Z
\left(
\left. \frac{\partial E_{\rm G} }{\partial v_{Z}}\right|_{\ve{v}=\ve{v}_0}
-\sum_l\Lambda_l\left.\frac{\partial g_{l} }{\partial v_{Z}}\right|_{\ve{v}=\ve{v}_0}
\right)^2
\end{equation} 
with respect to $\Lambda_l$.
Note that the lack of uniqueness for the  Lagrange parameters $\Lambda_l$
has no consequences for the fields~(\ref{sdfj}).
The latter are always uniquely defined, apart from a total 
energy shift that can be absorbed in the chemical potential. 

With the fields~(\ref{sdfj}) and the renormalization matrix determined,
we diagonalize~(\ref{tzs})
and determine $\ket{\Psi_0}$ by means of the standard tetrahedron method.

\bibliographystyle{unsrt}
\bibliography{bib4FG}

\end{document}